\documentclass[11pt,preprint]{aastex} 
\usepackage{graphicx,amssymb}
\usepackage{txfonts}
\usepackage{color}
\newcommand{\turck}{\textrm{Turck-Chi\`eze }}

\begin{document}
%
 \title{SEISMIC and DYNAMICAL SOLAR MODELS  \\
 I-THE IMPACT OF THE SOLAR ROTATION HISTORY ON \\NEUTRINOS AND SEISMIC INDICATORS}
   \author{S. Turck-Chi\`eze\altaffilmark{1},
   A. Palacios\altaffilmark{2,1}, J. P. Marques\altaffilmark{3},
	  P.A.P. Nghiem \altaffilmark{1} \footnote{P.A.P Nghiem has now left the Service d'Astrophysique in IRFU/CEA and works in another discipline in the same institut.}}

\shorttitle{The rotating Sun}
\shortauthors{S. Turck-Chi\`eze$^1$, A. Palacios$^{2,1}$, J. Marques$^3$, P. Nghiem}
 \affil{$^1$ CEA/DSM/IRFU/Service d'Astrophysique, CE Saclay, L'Orme des
              Merisiers - b\^at 709, F-91191 Gif-sur-Yvette (France)\\          
            $^2$ GRAAL, Universit\'e Montpellier II, CNRS, place Eug\`ene Bataillon, F-34095
              Montpellier (France)\\
             $^3$    Observatoire de Paris, 92 Meudon, (France)}
 \email{Sylvaine.\turck@cea.fr, Ana.Palacios@univ-montp2.fr, Joao.Marques@obspm.fr}
 \date{Version 1 Avril 2010}
 \begin{abstract}
Solar activity and helioseismology show the limitation of the
standard solar model and call for the inclusion of  dynamical processes
in both convective and radiative zones. In this paper, we concentrate on
the radiative zone. We first recall  the sensitivity of boron neutrinos to the microscopic physics included in solar standard and seismic models. 
We confront the neutrino predictions of the seismic
model to all the detected neutrino fluxes.  Then we compute new models of
the Sun including a detailed transport of angular momentum and chemicals due to internal rotation that includes meridional circulation and shear induced turbulence. We use two stellar evolution codes: CESAM and
STAREVOL to estimate the different terms.  We follow three temporal evolutions
of the internal rotation which differ by their initial conditions:  very slow, moderate and fast rotation, with magnetic braking at the arrival on the main sequence for the last
two. {bf We find}
that the meridional velocities in the present solar radiative zone  is extremely small in comparison with
 those of the convective zone (smaller than 10$^{-6}$ cm/s instead of m/s). All models lead to a radial differential rotation profile in the
radiative zone but with a significantly different contrast.
 We compare these profiles to the presumed solar internal rotation and
  show that  if meridional circulation and shear turbulence  were the
  only mechanisms transporting angular momentum within the Sun,  a rather slow
  rotation in the young Sun  is favored.
 We confirm the small influence of the transport by rotation on the sound speed profile but its
potential impact on the chemicals in the transition region between
radiation and convective zones. These models are physically more
representative of the real Sun than the standard or seismic solar models
but a high initial rotation, as it has been considered
previously, increases the disagreement with neutrinos and the sound speed in the radiative zone. This present work pushes us to pursue the inclusion of the
other dynamical processes  to better reproduce the observed solar profile in the whole radiative zone and  to better describe the young active Sun. We also need to get a better knowledge of solar gravity mode splittings to use their constraints.
  \end{abstract}

    \keywords{ hydrodynamics -- neutrinos-- Sun : evolution -- Sun: helioseismology --Sun : interior   --Sun:  rotation}

 \maketitle
%

\section{INTRODUCTION}
Evidence for the presence of dynamical processes exist in stars all
  over the Hertzsprung-Russell diagram. In the case of the Sun, rotation
  and magnetic activity have been confirmed for more than 4
  centuries. Rapid rotation is found mainly in young stars and intermediate-mass to
  massive stars while mild to slow rotation is found in low-mass stars and in giant stars.
 \cite{Kraft} was the first to study in details the projected rotational velocity $\rm v sin_i$ of young stars in clusters and shows the transition between rapid rotation for early-type stars and slow rotation for late-type stars. Then, Weber explained that the angular momentum loss is due to a magnetized wind in solar-like stars.   In studying Pleaides, Ursa  Major and Hyades,  \cite{Skumanich} used solar and cluster data to establish the law of variation of the rotation with age, this simple law is in fact more complex  and has been studied in great details for different masses \citep{Stauffer,Bouvier94,MaederMeynet00a,MaederMeynet04,Bouvier09}.
 
 Rotation affects the internal stellar structure and evolution both
  directly via the modification of the gravitational potential, and by
  means of associated transport processes.  It has a direct impact on the
  shape of stars, that can be directly probed with interferometric data for
  the strongly deformed fast rotating stars
  \citep{Domiciano03,McAlister05,vanBelle06,Kervella2006,Domiciano2008}. In
  the case of the Sun, the effect is small  as shown in \cite{Piau} but of considerable importance
  (Zahn 2009; Rozelot 2009) to check the influence of the deep core
  rotation and of the deep magnetic field \citep{Duez09, Duez10} on the tachocline and on the
  solar surface.

The standard solar and stellar evolution models do not take into
  account the effects of such dynamical processes as rotation or magnetic
  fields. If helioseismology required the improvement of the solar model
  and motivated the introduction of microscopic diffusion, going beyond the
  zero-order model, it is the far too rough agreement between observational
  data and theoretical predictions for other stars that has led to
  progressively introduce the dynamical physical processes likely to affect
  the transport of momentum and chemicals in the stellar evolution
  codes.  Using either a purely diffusive approach (\cite{Endal,
  Pinsonneault, Chaboyer, Langer, Heger00}) or the more complex formalism
  developed by \cite{Zahn92} and \cite{MaederZahn98}, the introduction of the
  rotation-induced transport of angular momentum and chemicals to model
  both massive and low-mass stars improves the comparison with observations
  \citep[see e.g.][]{MeynetMaeder00,TalonCharbonnel98,Palacios03}. These
  results encourage us to pursue this effort for the Sun to get a proper
  interpretation of the existing helioseismic observations and the coming
  asteroseismic ones.

Acoustic and gravity mode detections provide a unique insight on the
internal solar sound speed, density and rotation profiles
\citep{Kosovichev,Turck2001a,Thompson,Couvidat,Turck04a,Mathur,Mathur2}.
This gives us a unique opportunity to validate the complex formalism
introduced in rotating stellar evolution models.  This information
justifies the development of a Dynamical Solar Model to be confronted with
the helioseismic and neutrino probes. The solar status must improve our
knowledge of solar-like stars where only external stellar rotation rates or
abundance anomalies are used to confront theoretical assumptions to
observational facts. Moreover, building a consistent dynamical evolution
model for the Sun will largely contribute to establish a complete and
consistent MHD representation of the Sun and a good connection between
internal and external magnetism.

The inner solar rotation is the first evidence of the internal
dynamics that needs to be understood. This is not an easy task because
the present-day profile results from the interplay between several
distinct processes
\citep{ZahnTalonMatias97,Chaboyer,Eggenberger05,CharbonnelTalon05} and
their influence depends on the adopted theoretical prescriptions.  The
most recent of these studies include the transport of angular momentum
by magnetic field  or internal gravity waves, in addition to the
``purely'' rotational transport by meridional circulation and
shear-induced turbulence, to obtain a more efficient extraction of
angular momentum and a relative flat rotation profile in the radiative
interior in better agreement with helioseismic data.  Actually, both
magnetic field and internal gravity waves probably contribute to the
transport of angular momentum in stellar interiors, yet no models have
been published up to now including all the processes.
Moreover, the way to account for the effect of magnetic fields in 1--D stellar evolution
codes is still puzzling and a matter of debate, and the introduction of gravity waves produces some
irregularities in the radiative rotation profile \citep[see e.g.][]{CharbonnelTalon05} in disagreement with
the helioseismic results.

 In the present paper, we limit ourselves to the sole effect of
 rotational transport and carefully study the hypotheses and the order of magnitude of the terms that we
 introduce for the inner rotation. This approach pushes further
 the previous works by
 \citet{ChaboyerZahn92,Chaboyer,CharbonnelTalon05,Eggenberger05,Yang1,Yang2}.
We focus our analysis on the solar core and we compare our results with all the recent
 existing seismic and neutrino indicators.  This paper is a first of a
 series where we will discuss in details the influence of rotation,
 magnetic field and internal waves on the solar dynamical model.  We do not  present and discuss  the abundances of lithium and beryllium as it has been done extensively by \cite{Pinsonneault}. Indeed we have shown in our previous works on the tachocline \citep{Brun2} and on young stars that  the abundance of these elements are sensitive to the early evolution as well as to the presence of the tachocline.  These studies lead us to the conclusion that we need to treat properly the magnetic field in the early phase to better estimate the lithium evolution during this phase \citep{Piau}.

In Section 2, we present the status of the solar classical models and
compare the seismic model predictions to all the present detected
neutrinos. Then in Section 3 we recall the formalism proposed by
\citet{Zahn92} and \citet{MaederZahn98}, and slightly improved by
\citet{Mathis1} that we use to introduce the rotation effects in the
stellar evolution codes. In that section, we also give a brief
description of the CESAM and STAREVOL stellar evolution codes used in
the paper, and present the three different types of solar models which
differ by their initial rotation and the presence of magnetic braking.
In Section 4 we compare the results of these three rotating solar
models obtained with both codes. These models are compared with
seismic observations and neutrino detections in Section 5.  In the
last section, we summarize the important points in a more general
context.

\section{THE MICROSCOPIC PROCESSES AND THE SOLAR PROBES}

The Sun differs from other stars by its proximity and our 
  ability of observation. We know better its
luminosity, radius and mass than for any other star. This has, for long, turned the Sun
  into a reference for stellar evolution. Nevertheless  the
  numerous constraints point to flagrant 
limitations of the standard picture:
\begin{itemize}
\item
In this framework, the luminosity increases by 30\% in 4.6 Gyr, 8\%
during the last Gyr and by only 10$^{-8}$ during the last century. But
in fact, the so-called solar constant shows variations by about
$10^{-3 }$ in mean value \citep{Frohlich06} correlated to the solar
cycles (ACRIM and SoHO observations). The presence of faculae around
the sunspots explains the increase of luminosity at the maximum of the
cycle.  Moreover the series of data shows clearly short time scale
luminosity variation connected to the rotation and the presence or
absence of sunspots in the field of view.  This fact shows, just from
fundamental quantities, that the standard framework is not sufficient
and that magnetism and rotation must be introduced in the solar
structural equations to interpret these observations because it is
believed that these manifestations have an internal origin
\citep{Turcklambert,Duez09}.
\item
We have learned recently that the absolute irradiance is not so well established. The SORCE satellite
suggests a reduction of nearly 6 W/m$^2$ \citep{Kopp}  in
comparison with the standard ($3.846 \pm 0.004 10^{33}$ ergs/s
corresponding to 1367.6 W/m$^2$) used for solar model,  see Bahcall \& Pinsonneault (1995). This result modifies
by 1.5-2 W/m$^2$ (340 instead 342 W/m$^2$) the energy which reaches the
stratosphere. So, the present solar luminosity value, which is used to
calibrate the standard model,  is presently uncertain at 0.5\% \citep{TurckLefebvre}.
The calibration and ageing of the radiometers are difficult to estimate, so the use of the space data is not easy. The SORCE/TIM instrument has an estimated
absolute accuracy of 350 ppm and a long-term repeatability of 10 ppm per
year. Nevertheless, the analysis of the 30 years data series seems to show
a slow decrease of the total luminosity \citep{Frohlich06} between
successive minima which cannot be  
understood in the standard solar
framework.

\item
 For three decades, our star is also being scrutinized by two probes of the
interior (neutrinos and seismic modes) that help us to largely progress on
our capability to check the internal solar plasma in great details. We are
now able to disentangle the production of neutrinos for specific nuclear
reactions like $\rm^7Be (p, \gamma)\; and \; ^8B (p, \gamma)$, and to sum
the different flavours to really estimate the number of
emitted neutrinos. 
The quality of the detection of the boron neutrinos (see \S~2.2 and
  Tab.~\ref{tab:neutri}), is such that the central temperature of the Sun
  is now known to about 0.5\%. Neutrinos can thus bring very strong constraints to the
  solar models.
On the helioseismic side, the
detection of million acoustic modes stimulated a real insight on the
thermodynamics of the radiative region (see below) and on the dynamics of
the convective zone. Gravity modes appear also promising to reveal the last
missing information, the dynamics of the  solar core \citep{Turck04b,Garcia2007, Garcia08,Mathur,
Mathur2}.
\end{itemize}

\subsection{The validity of the standard solar model}

The solar radiative zone represents 98\% of the total mass of the Sun and
the adopted microphysics plays a basic role in this region.  The
equilibrium between gravitational energy, nuclear energy production and the
energy escaping by photon interaction needs to be followed in the radiative
interior over long time-scales.  This region is now permanently probed by
helioseismology, which is a key to validate the various ingredients used in
the construction of the standard solar model.  One success of the SoHO
space observatory has been obtained by measuring the Doppler velocity
shifts down to the low frequency range of the acoustic spectrum with two
instruments: GOLF, Global Oscillations at Low Frequency described by
\cite{Gabriel} and MDI, a Michelson Doppler Imager built by
\cite{Scherrer}. This part of the spectrum, contrary to the high frequency
range, is not sensitive to the variability of the sub surface layers along
the solar cycle \citep{Turck2001a, Couvidat}. The corresponding modes have
longer lifetimes but smaller intensities \citep{Bertello00,Garcia01}. From
these observations,  a very clean sound speed profile has been established down to $0.06
~R_\odot$ together with a reasonable density profile. It has been compared 
to the theoretical models for different solar compositions. For the latest
sets of solar abundances with lowered oxygen abundance \citep{Holweger01,Asplund2009,Caffau08}, the theoretical
profiles show clear differences compared to the observed soundspeed, that
still need to be understood \citep{Turck2001a,Turck04a}.

\subsubsection{The nuclear processes}

It is interesting to notice that each physical ingredient of the structural
  equations (specific nuclear rate, specific opacity coefficient, screening
  or Maxwellian tail distribution) has a specific signature on the sound
  speed profile \citep{Turck97}. It has thus been of prime
  importance to check the validity of the involved nuclear processes.
  \cite{Turck2001b} note that the present sound speed profile does
  not favour any tiny variation of the Maxwellian distribution of the ions
  nor strong screening or large mixing in the core. The core sound speed
  profile and the specific signature of the pp reaction rate put strong
  observational constraint on the cross section at a level of 1\% including
  the screening effect. This cross section is known only theoretically due
  to the weak character of the interaction. The other nuclear cross
  sections have been measured in laboratory during the last three decades
  but the extrapolation towards the stellar plasma conditions has been only
  measured for the $\rm (^3He, ^3He$) interaction.
  
   At present, the
  reaction rates are generally considered to be reasonably well under
  control, and in the CESAM code in particular, we use the most recent
  estimates for the $\rm^7Be(p, \gamma)^8B$ \citep{Junghans}\footnote{The
  astrophysical S-factor determined by \cite{Junghans} is S$ _{17}(0)= 21.4
  \pm 0.5 \,\rm(stat) \pm 0.6 \,(syst)$. Some slightly higher value is
  still discussed by \citet{Igamov}} and $\rm^{14}N(p, \gamma)^{15}O$
  \citep{Formicola}. That last cross section leads to a consequent
  reduction of the CNO contribution to the nuclear energy in main sequence.
  Some uncertainties could remain on the screening effect of CNO processes
  but they play a rather small role in the present confrontation of the
  models to the two mentioned probes.  Tables~\ref{tab:neutri}
  and~\ref{tab:neutrisis} summarize  our previous and present works. For a comparison between the values of table 1  and the work of Bahcall and collaborators along years, the reader is refered to \cite{Turck04n}. Table 2 takes into account the neutrino oscillations determined by  $\rm \Delta m^2_{12}= 7 \; 10^{-5} eV^2$ and $\rm tg^2 \theta_{12}= 0.45$ recommended by \cite{BahcallGaray}, these values are compatible with the most recent estimates within the error bars. 

\subsubsection{The impact of the solar composition on the radiative zone properties}

One important question has emerged from the new spectroscopic estimates of
the carbon, nitrogen and oxygen atmospheric abundances, leading to a
reduction by 20-30 \% of the related abundances: {{\em" Are we so sure that we
understand properly the inner composition of the Sun in the radiative zone"
?} This question has an important issue because the solar abundance is
choosen as a reference for the stellar evolution models. For instance, up to
recently, the abundance of these nuclides was larger than that of any other
heavy element by about a factor 10 in Population II stars, so the change
in the solar CNO abundance has a strong impact on the evaluation of their
chemical pattern \citep{Turck04a}. The situation is now
clarified because the three independent analyses
\citep{Holweger01,Asplund2009,Caffau08} converge to about the same
reduction. A reduction by a similar factor (30\%) of the iron content also appeared in
the early nineties, but with rather small effects on the whole radiative
zone \citep{TurckLopes93}.

\begin{figure}
\includegraphics[width=70mm, angle= 90]{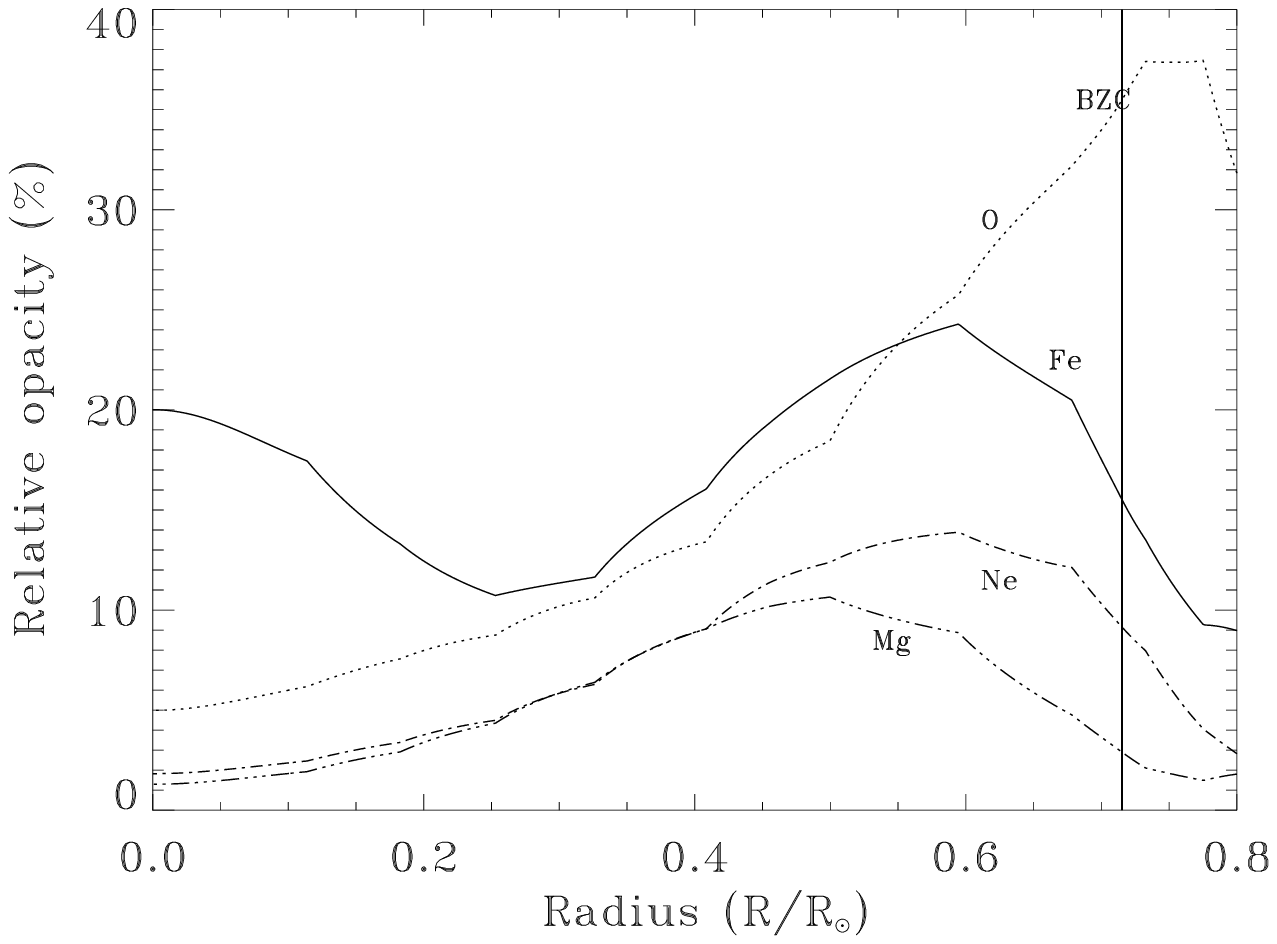}
\includegraphics[width=70mm, angle= 90]{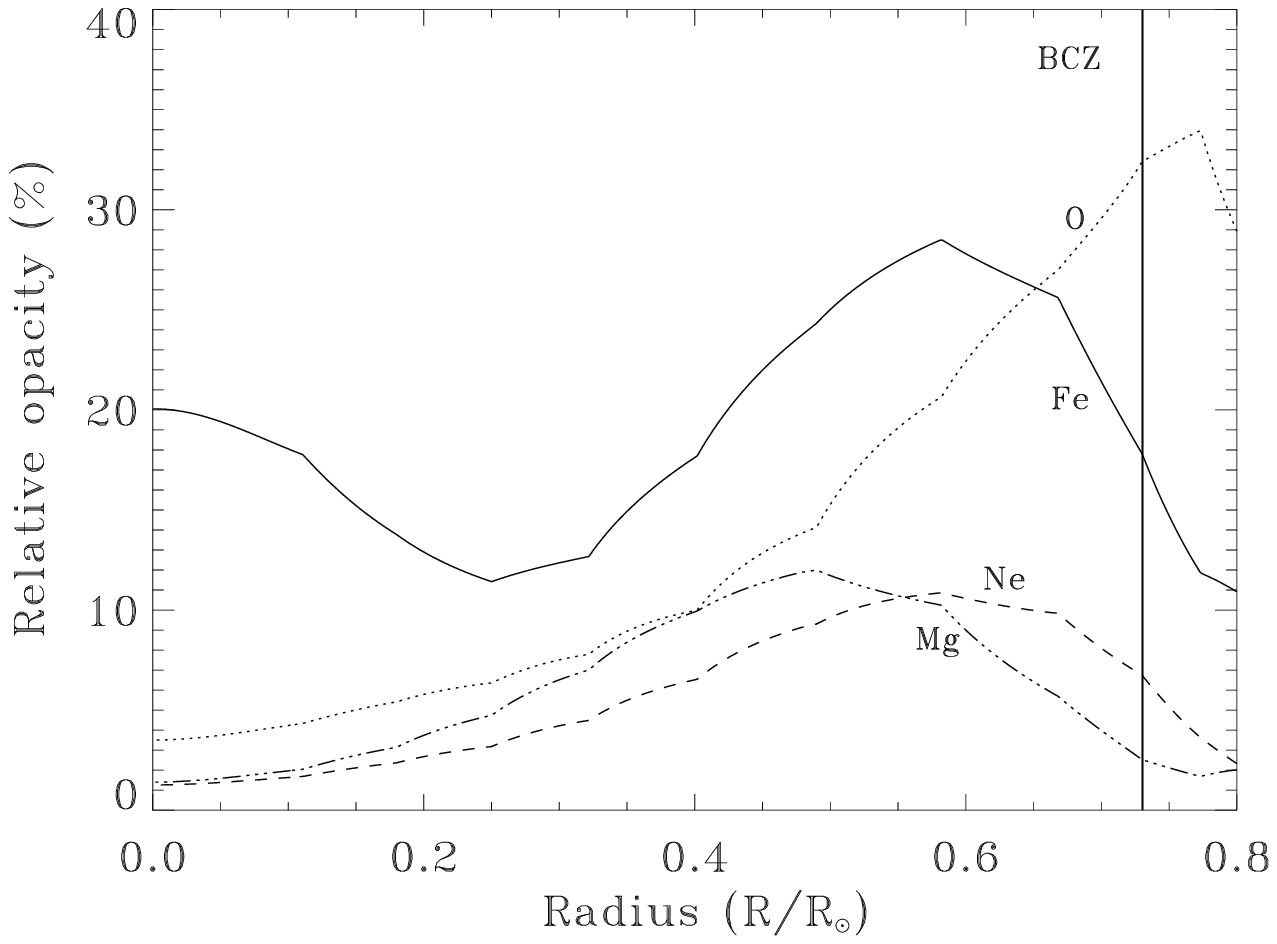}
\caption{ \small Relative contribution of the most important heavy element
contributors to the total opacity when one considers the solar composition
proposed by  Grevesse \& Noels (1993) ({\em left}) and Asplund et al. (2004) ({\em right}) . An intermediate value of
the oxygen contribution is now more compatible with the recent studies
\citep{Asplund2009,Caffau08,Caffau09}. We have noticed that slight
differences exist between  the different opacity tables due to slight
different states of ionization of the different species
\citep{Turck2009}.
\normalsize}
\label{fig:figure1} 
\end{figure}

As far as CNO elements are concerned, this substantial reduction has a real impact on
the radiative sound speed profile mainly due to the subsequent change in
the oxygen opacity (see Fig.~\ref{fig:figure1}) and modifies the depth of
the convective envelope \citep{Turck04a,Bahcall05}. }
The confrontation between helioseismic indicators and results of standard
solar models was examined extensively (\turck et al. 2004; Bahcall et
al. 2005, Guzik et al. 2005). It is frustrating to note that the sound
speed and density profiles of the model including the updated solar
chemical composition are very similar to those corresponding to the case
where we omitted to account for the slow gravitational settling of the
nuclides \citep{Thoul94,Brun} and that this last update destroys the
apparent agreement with the observed one. 
We may thus quantify the effect of the change of composition since the inclusion of
the microscopic processes corresponds to a reduction by 10-12\% of
all the heavy elements. Consequently a 20\% variation of CNO
abundances is almost equivalent to the 7-9 \% variation of all the
heavy elements due to gravitational settling, or to an equivalent change of
opacity coefficients if all the elements were concerned. See \cite{Turck97}
where the impact of a lot of changes are considered and
Table~\ref{tab:neutri} for the
consequence on neutrino predictions.

Effectively, a change in the CNO abundances also affects the mean total
opacity of the stellar plasma, and thus the evolution of the Sun up to its
present status as well as its present structure.  The knowledge of the
opacity coefficients used in a solar model is up to now purely theoretical.
The conditions of temperature and density in the radiative zone ensure that
the plasma is only fully ionized for its main constituents, hydrogen and
helium, but heavier species such as iron down to oxygen become partially
ionized in the radiative zone as summarized in Fig.~\ref{fig:figure1}. It
is why the bound-bound and the bound-free interaction of photons with
matter limit the evacuation of the energy produced within the inner 25\% of
the radius, corresponding to half the solar mass
\citep{IglesiasRogers96}. The small amount of iron (only 2.8 10$^{-5}$ of
the hydrogen contribution in fraction number) contributes to about one
fifth of the opacity cross section in central conditions (see
Fig.~\ref{fig:figure1} and \cite{tur93} for more details). The oxygen
is the second contributor and plays a crucial role to determine the base of
the convective zone.

Up to now the details of the ion interactions have been verified only
indirectly through acoustic pulsation eigenmodes. These modes probe plasma
properties throughout the Sun from 6\% R$_\odot$ up to the limit of the
convective zone. However they depend on both the detailed composition and
opacity. To disentangle both contributions would require a high radial
resolution of the seismic data allowing to capture some specific
bound-bound effect through bumps in the radial sound speed profile in the
radiative zone \citep{Turck98}, that is still out of reach at present. 

Some proposed
solutions to reconcile the standard model with revised CNO abundances and
helioseismology \citep{Guzik,Lin07,Basu08} have not been retained.  Some others seem to favor  the old composition \citep{Basu07, Serenelli09}. We explore, here, the idea that the
   standard solar model (SSM) framework is questionable.   Considering that any improvement has always been considered like a new informative fact 
   and that a better determination of CNO was waited since the nineties \citep{tur93}, the present
  facts suggest three reasonable solutions which could all exist simultaneously:
\begin{itemize} 
 \item 
 The gravitational settling is not well constrained today for CNO and heavy
 elements.  The photospheric helium content
 deduced from helioseismology \citep{Vorontsov} of $\rm Y = 0.25 \pm 0.03$, and confirmed by \cite{Basu95}  for OPAL equation of state has demonstrated the need to  take into account this slow
 atomic diffusion introduced first by \cite{Cox89}. This
 process leads today to a reduction of  about 10-15\% of the He
 mass fraction at the solar surface \citep{JCD93,Thoul94,Brun,Michaud04} but we cannot check
 yet if the order of magnitude is correct also for CNO and
 iron.
\item
The opacity coefficients used in the models have never been verified by
  laboratory measurements under the physical conditions of the stellar
  interiors and may be a source of uncertainty.  The high energy lasers
  developed in France (LIL, LMJ) and in the United States (NIF) will offer
  conditions corresponding to the solar radiative zone and might help
  constraining these quantities.  We are preparing experiments to measure
  the absorption energy spectrum of isolated elements r mixtures in plasma
  conditions which are more and more useful for the stellar community
  \citep{Bailey, Loisel,Turck2009}. Laboratory opacity and equation of
  state experiments might benefit in the coming decade from an equivalent
  effort to the one dedicated to nuclear reaction rate measurements.
 A detailed verification of the interaction between photons and plasma will
contribute to disentangle the different processes in the deep interior of stars. 

\item 
The radiative zone of the Sun is not well reproduced by the hypotheses
of the SSM, and we need to improve the models by the addition of
  dynamical processes. It is one of the reasons to develop the Dynamical Solar Model (DSM)
which must introduce the dynamics of the tachocline, the gravity waves
generated at the base of the convective envelope and propagating
  in the radiative interior, the presence of a potential magnetic
field and the different transport processes induced by the rotation.    \cite{Guzik10} examine the idea of mass loss and accretion, and in a coming work, we question the hypothesis of the energetic balance  \citep{Turck10}.
 \end{itemize}

Such a complete model of the Sun is not ready to be compared to seismic and
neutrino probes, it is why we have developed an intermediate model, the
seismic model described in the following section. A first step toward the
DSM is then to confront all the probes with solar models including
rotation, which is the purpose of the present paper. We shall produce
models in the following sections which can be considered as the DSM1
step.  \small
\begin{table}
  \begin{center} 
    \caption{Time evolution of the boron neutrino flux prediction (expressed in 10$^6 /cm^2/s$) associated with the 
    reaction $^7{\rm Be} \,(p,\gamma) \,^8{\rm B} \longrightarrow  \,^8{\rm Be}* +e^+ 
    +\nu_e \longrightarrow  \,^4{\rm He}$, for 
calibrated    solar structural models. Also mentioned are the corresponding central 
    temperature $T_c$, initial helium content $Y_0$ and the origin of the improvements 
    introduced in the corresponding solar model. }
\vspace{1em}
    \renewcommand{\arraystretch}{1.2}
    \begin{tabular}[h]{lccccc}
  \hline  
Year	& Boron  neutrino flux    &	$T_c$	& 		$Y_0$       &    	Problem solved	\\	
1988  Turck-Chi\`eze et al. & $3.8 \pm 1.1$      &    15.6      &  	      0.276   &       SSM CNO opacity, $^7{\rm Be}(p, \gamma)$ \\
1993  Turck-Chi\`eze \& Lopes & $4.4 \pm 1.1$      &    15.43     &          0.271  &       SSM Fe opacity, screening \\
1998  Brun et al. & 4.82       &          15.67  &             0.273 &        SSM Microscopic diffusion	\\
1999  Brun et al. & 4.82        &         15.68   &            0.272  &       SSM Turbulence in tachocline \\
2001 Turck-Chi\`eze et al.  & $4.98 \pm 0.73$  &    15.74        &       0.276      &   SSeM \\
2003 Couvidat et al.  & $5.07 \pm 0.76$    &   15.75         &      0.277     &   SSeM +magnetic field \\
2004 Turck-Chi\`eze et al. & $ 3.91\pm 1.1$    &   15.54          &     0.262    &     SSM Asplund composition 	\\
2004 Turck-Chi\`eze et al. & $ 3.98\pm 1.1$    &   15.54          &     0.262    &     SSM Holweger  composition 	\\

2004 Turck-Chi\`eze et al.  & $5.31 \pm 0.6$    &     15.75       &        0.277      &   SSeM $\& \rm \,new\, ^7{\rm Be}(p,
\gamma)$, $^{14}{\rm N}(p, \gamma)$ 	\\
Present work & 5.31 $\pm$ 0.6 & 15.75  & 0.277  & SSeM &  \\
&  4.21   &  15.51 & 0.262  & SSM Asplund 2009 \\
			&  5.09     &  15.64  & 0.273 &  SSM GN and DSM1 GN
with slow rotation  \\
			& 4.52      &    15.54  & 0.269 & DSM1 GN moderate rotation\\ 			
 \hline
   \end{tabular}
    \label{tab:neutri}
    \end{center}
\end{table}
\begin{table}
  \begin{center} 
    \caption{Comparison of the calibrated Solar Seismic Model predictions and the  detection of the different neutrinos.}
  \vspace{1em}    
    \begin{tabular}[h]{c}
     \hline
 SNO predictions $5.31 \pm 0.6 $ $10^{6}$ $\rm cm^{-2}$ $s^{-1}$ \\
SNO results  $5.21 \pm 0.27 \,({\rm stat})\, \pm 0.38 \, ({\rm syst})$ 10$^{6}$cm$^{-2}$s$^{-1}$\,  \\
 \hline
 Gallium prediction without neutrino oscillations: $123.4 \pm 8.2$ SNU \\  
 Gallium prediction with neutrino oscillations: $66.65 \pm 4.4$ SNU \\  
 Gallium detection:  GNO+Gallex: 70.8$\pm$  4.5$\pm$ 3.8 SNU; SAGE:  70.8$\pm$  5.3$\pm$ 3.7 SNU\\
  \hline
 Chlorine prediction without neutrino oscillations $7.6 \pm 1.1$ SNU  \\
 Chlorine prediction with neutrino oscillations: $2.56 \pm 0.23$ SNU  \\ 
 Chlorine detection:  $2.56 \pm 0.16 (\rm stat) \pm 0.16 (\rm syst) $ SNU  \\ 
  \hline
$^7Be$ prediction without neutrino oscillations: 4.72 10$^{9}$cm$^{-2}$s$^{-1}$  \\
$^7Be$ prediction with neutrino oscillations: 3.045 10$^{9}$cm$^{-2}$s$^{-1}$\\
BOREXINO detection 49 $\pm$ 3 (\rm stat) $\pm$ 4 (\rm sys)  cts/day/100 tons or  3.36 $\pm$ 0.365 10$^{9}$cm$^{-2}$s$^{-1}$\\
\hline
\end{tabular}
 \label{tab:neutrisis}
    \end{center}
\end{table}
\normalsize

\begin{table}
\center
\caption{Characteristics of the rotating solar models computed with
  CESAM (A$_C$, B$_C$) and STAREVOL (A$_S$, A'$_S$, B$_S$, C$_S$)
  codes. All STAREVOL models where computed using $\alpha_{MLT} = 1.7378$
  and initial helium mass fraction $Y_{ini} = 0.280.$  In this study, all the CESAM models are calibrated in luminosity and radius at the present age.}
  \vspace{0.5cm}
 \begin{tabular}[H]{ l c| c| c| c| c|}
\hline
&Model & D$_h$ & J$_{initial}$ & $\upsilon_{ZAMS}$ & $\upsilon_{\odot}$\\
\hline \hline
&A$_C$ & Mathis et al. (2004) & 3.27 10$^{48}$ & 2.19 km.s$^{-1}$ & 2.13 km.s$^{-1}$ \\
&A$_S$ & Mathis et al. (2004) & 4.84 10$^{48}$ & 2.15 km.s$^{-1}$ & 2.05km.s$^{-1}$\\
&A'$_{S}$ & Zahn (1992) & 4.84 10$^{48}$ & 2.15 km.s$^{-1}$ & 2.03km.s$^{-1}$ \\
&B$_C$ & Mathis et al. (2004) & 1.10 10$^{49}$ &  19.6 km.s$^{-1}$ & 2.09 km.s$^{-1}$\\
&B$_S$ & Mathis et al. (2004) & 3.88 10$^{49}$ & 19.7 km.s$^{-1}$ & 3.08km.s$^{-1}$ \\
&C$_S$ & Mathis et al. (2004) & 8.74 10$^{49}$& 53.2 km.s$^{-1}$ & 3.03km.s$^{-1}$ \\\hline
\end{tabular}
\label{tab:models}
\end{table}

\subsection{Prediction of neutrino fluxes, gravity mode frequencies and structural profiles from SSeM}

The quality of the seismic observations is such that one can build a
seismic model. This model is designed to reproduce the measured solar sound
speed using the classical stellar evolution equations and slightly changing
some recommended specific absolute values of physical inputs within their
own uncertainties \citep{Turck2001a,Turck04a,Couvidat03a}. It can be
established for the new solar composition and will converge to the same
predictions by definition.

The interest of such a model comes from the idea that the standard
 framework is obviously too crude to reproduce all the existing observed
 quantities. This model is useful to improve the
 prediction of the neutrino fluxes (Tables~\ref{tab:neutri} and ~\ref{tab:neutrisis}) or the gravity mode
 frequencies because it implicitly includes the known sound speed profile
 \citep{Turck04a,Couvidat03a,Mathur}.

Table~~\ref{tab:neutri} illustrates the time evolution of the predicted $^8B$ neutrino
flux which depends strongly on the central temperature and
consequently, on the details of the plasma properties. The seismic
model prediction  \citep{Turck01,Couvidat03} agrees remarkably well with the measured value
obtained with the SNO detector (filled with heavy water), which
  is sensitive to all the neutrino flavours. For the gallium or water
detector predictions, one needs to inject the energy dependent
reduction factor due to the fact that the electronic neutrinos are
partially transformed into muon or tau neutrinos. This property of neutrinos has been confirmed 
last year by the Borexino results \citep{borexino}. Doing so, the
agreement between the predictions of the seismic model and all the
detectors is extremely good \citep{Turck05a}. Table 2 confirms that it is also true for the beryllium neutrinos.

Another interesting property of the SSeM is the fundamental periodicity
$P_0$ which characterizes the asymptotic behaviour of the gravity modes
nearly equally spaced in period for frequencies below $\sim 100 ~\mu{\rm
Hz}$.  $P_0=2{\pi}^2 \left(\int_{0}^{r_c} \frac{N}{r} dr \right)^{-1}$
where $N$ is the Brunt-V\"ais\"al\"a frequency, $r_{c}$ is the radius at
the convective bottom. Before the launch of SoHO, there was a great
dispersion on the theoretical predicted values of $P_0$.  Following
\cite{Hill91}, its value was varying from 29 minutes to 63 minutes
depending on the models. Today, standard and seismic models agree within 1
minute, which helps to predict the general properties of the
gravity modes \citep{Mathur} because models have been improved by the understanding of the property of the modes and their sensitivity.
Nevertheless  the present standard model, including the new CNO composition, largely deviates from the seismic observations, consequently the value of  $P_0$ might be better determined by the seismic model or any coming model in agreement with seismic observations. Of course the detection of individual gravity modes might improved the density profile in the core and consequently will put more pressure on the seismic model in the core.

Despite the remarkable agreement between the two probes of the central
region of the Sun (neutrinos and helioseismology), SSeM is not a physical
model and  SSM predictions with the new solar composition encounter a series of difficulties.  In the following section,
we present a first step to transform SSM or SSeM into a more realistic
DSM1, by introducing rotation and the associated transport of matter and
angular momentum in the computations.

\section{ HOW TO MODEL THE INTERNAL ROTATION IN 1D STELLAR EVOLUTION CODES} 

\subsection{{\bf Meridional circulation and shear-induced turbulence :
    adopted formalism }}

The present Sun is a slowly rotating star ($\simeq$ 2.2
  km/s at the equator, 1.7 km/s near the poles).
It is thus a reasonable first order approximation to treat rotation with a perturbative
  approach and to adopt the hypothesis of spherical symmetry in the standard model framework.
We may nevertheless recall that the solar superficial deformations can be measured and that
the quadrupolar moment is of the order of 1.84 $10^{-7}$ \citep{Lydon} and will be improved by the microsatellite
PICARD \citep{Thuillier05}. 

Even if departures from spherical symmetry can be considered small in
the solar case, they are sufficient for rotation to induce large-scale circulations in the
radiative and convective interior, that will simultaneously advect
angular momentum, and chemical species
\citep{Busse82,Zahn92}. Moreover, in case of radial differential
rotation, which could be the case in the solar deepest interior,
different hydrodynamical instabilities may develop that will generate
hydrodynamical turbulence in the radiative regions. Laboratory
Couette-Taylor experiments conducted for Reynolds numbers $Re \simeq
10^6$ by \cite{RichardZahn99} indicate the development of turbulence
in the flows. This gives a good hint that in stellar interiors, where
the typical Reynolds number is of about $10^{12}$ (solar case), the
hydrodynamical instabilities associated with differential rotation
eventually become turbulent.

 Following \citet{Zahn92,MaederZahn98,Mathis1}, the transport of angular
   momentum and chemical species by meridional circulation and
   shear-induced turbulence has been added to the stellar structure
   equations in order to estimate the effects of such a transport on the
   evolution and the structure of the Sun up to  the present time.
   
Under the assumption of shellular rotation ensured by a strong anisotropy
   of the turbulent diffusivities $D_h >> D_v$, all the relevant variables
   and vectorial fields may be described as the sum of the mean value over
   an isobar and of a second (or fourth) order perturbation
   \citep{Zahn92,Mathis1}. This means that the variables are projected on a
   basis of Legendre polynomials, an approach very similar to that used in
   helioseismology.  This allows to separate angular and radial parts and
   thus to account for the rotational transport of angular momentum and
   chemicals over secular timescales in 1--D stellar evolution codes, while
   2--D or 3--D secular stellar evolution is yet far from maturity.

Using this formalism, the momentum equation

\begin{equation}
\rho \left[\frac{\partial \vec{V}}{\partial t} + \left ( \vec{V} . \vec{\nabla} \right) \vec{V} \right] = - \vec{\nabla}P - \vec{\nabla}\phi + \vec{\nabla} . || \tau ||
\end{equation} 
\noindent
becomes, when averaging over the isobar and using an azimuthal projection :

\begin{equation} 
 \rho \frac{d}{dt}{ \left (r^2  \overline{\Omega} \right )} = \frac{1}{5 r^2}\frac{\partial}{\partial r} \left(\rho r^4 \overline{\Omega}U_2 \right) + \frac{1}{r^2}\frac{\partial}{\partial r} \left(\rho \nu_v r^4 \frac{\partial \overline{\Omega}}{\partial r} \right)
\label{eq:AM}
\end{equation}
where $\overline{\Omega} = \Omega(r)$ is the mean angular velocity on the
isobar of radius $r$ and $ \nu_v$ is the vertical turbulent viscosity associated to the shear
instability (see Eq.~\ref{eq:Dv} below).\\
$U_2$ is the vertical meridional velocity component :
\begin{equation} 
 U_2 = \frac{P}{\rho g C_p T [\nabla_{\rm ad} - \nabla +(\phi/\delta)
 \nabla_\mu]} \left(\frac{L}{M_\star}(E_\Omega + E_\mu)\right),
\label{eq:U2}
\end{equation}

\noindent with P is the pressure, $\rm {C_p}$ the specific heat, E$_\Omega$
 and E$_\mu$ are terms depending on the 
$\Omega$- and $\mu$- distributions respectively, up to
the third order derivatives and on various thermodynamic quantities
\citep[see details in ][]{MaederZahn98}.

 Equation~\ref{eq:AM} is split into a system of four first-order equations in $\Omega$ complemented
by an equation describing the evolution of the horizontal mean molecular weight
fluctuations due to the vertical $\mu$ gradient for a given horizontal
turbulence and a vertical velocity:

\begin{equation} 
 \frac{d \Lambda}{dt} + U_2(r) \frac{\partial ln \overline{\mu}}{\partial r}= -6 D_h \Lambda  
\end{equation} 
where $\Lambda = \frac{\tilde{\mu}}{\mu}$, and $D_h$ is the horizontal
turbulent diffusivity.\\

The equation for the transport of nuclides by meridional circulation and
shear-induced turbulence can be written as a diffusion equation \citep[see
e.g.][]{ChaboyerZahn92} :

\begin{eqnarray}
  \left(\frac{{\rm d} X_i}{{\rm d}t}\right)_{M_r}= \frac{\partial}{\partial
  M_r}\left[(4\pi
  r^2\rho)^2\left(D_v+D_{\rm eff}\right) \frac{\partial X_i}{\partial M_r}\right] + \left(\frac{{\rm d}X_i}{{\rm d}t}\right)_{\rm nucl} + \left(\frac{{\rm d}X_i}{{\rm d}t}\right)_{\rm micro},
\end{eqnarray}
\noindent
where $X_i$ is the mass fraction of the i$^{th}$ nuclide, the second and
third terms are respectively the nuclear (nucl) and gravitational settling
(micro) terms.
$D_v \equiv \nu_v$ is the vertical component of the turbulent diffusivity. $D_{\rm eff}$ is the diffusion coefficient associated to the action of the meridional circulation, and it is given by 
\begin{eqnarray}
D_{\rm eff} = \frac{\left( r U_2\right ) ^2}{30D_h}.  
\end{eqnarray}

 These transport equations have to be solved  at each evolutionary
 timestep, adding a set of five coupled non-linear equations to the usual set of structure equations.
 This explains the small number of
 stellar evolution codes including this treatment: the STAREVOL code,
 \citep{PCTS06} and the Geneva stellar evolution code
 \citep{Eggenberger08}. It  has been recently included in the
 CESAM stellar evolution code.

Following the effects of rotation according to the above described
formalism is even more difficult in the solar case due to the small
perturbation they cause on the solar structure. In order to test the
quality of our results, we have decided to confront the results obtained
with two different stellar evolution codes using distinct numerical
approaches. We have computed rotating solar models with both STAREVOL
and CESAM codes. In the following, we present the codes, the numerical
implementation of the Maeder \& Zahn formalism and the main characteristics
of the solar models that will be discussed in the forthcoming sections.
 
\subsection{{\bf Stellar evolution codes}}\label{sec:codes}

\subsubsection{{\bf CESAM}}
The CESAM code has been developed by a consortium of french
astrophysicists to get a stellar evolution code of high accuracy for
numerous seismic uses. This code is robust and used by a large
international community. It is free of access for the basic use
(http://www.oca.eu/cesam), that means for standard physics excluding
the dynamical processes. Details of the code are described in
\cite{Morel}. This code does not use the natural variables to avoid
the singularities at the center. Moreover CESAM does not solve the non
linear problem of the limited conditions in the physical space but in
the B-splines space. This allows a good continuity of the functions and
their derivatives. In fact, this code has been especially built for an
helio and asteroseismic perspective, so all the variables used to
calculate the frequencies of the acoustic and gravity modes are
correctly calculated. The treatment of the transport equations by
rotation has been recently improved by Marques (2010), it is now done in the physical space with a projection of the
quantities, associated to the rotation, on the vectorial spherical
harmonics. These equations are solved independently of the structural
equations, and then all the interpolations from one model to the next
one (in time) are done on the B-spline basis.

CESAM is used by our team since 1995 (see the references in Table 1), in
 replacement of the Paczynski BINARY code used in the eighty for the
 predictions of helio and asteroseismic probes and for neutrinos.  With
 time, the code has been enriched by the different updated pp and CNO cross
 sections of the hydrogen burning. For the Sun, we use the recommended
 values of the Seattle meeting \citep{Adelberger}, then the most recent
 updated quoted in section 2. We have included the appropriate screening
 and use the Mitler prescription \citep{Dzitko} and the microscopic
 diffusion described in \cite{Brun}. CESAM is regularly updated with the
 most recent equation of state and opacity coefficients
 \citep{IglesiasRogers96, IglesiasRose96}. It included the different
 updates on the solar composition with an adapted low temperature opacity
 table \citep{Turck04a}. The models of this paper use the Holf atmosphere
  (see the difference between Kurucz and Holf atmosphere in
 \cite{Couvidat03a}).

For a comparison with the STAREVOL code, we have used the Grevesse and
Noels (1993) composition. This choice has no incidence on the results
because we show mainly relative comparison and because the solar structure is
only slightly modified by the effects of rotation.

\subsubsection{{\bf STAREVOL}}
The code STAREVOL has been applied to model stars all over the
  Hertzsprung-Russell diagram, and more specifically to study the effect of
  dynamical processes on stellar evolution and nucleosynthesis
\citep{Palacios03,CharbonnelTalon05,Decressin05,PCTS06,Decressin09}.
  Let us briefly recall here the main ingredients
  used for the computation of the solar model.  
We adopt the \cite{GrevesseNoels93} as reference for the solar chemical composition. The nuclear network includes 52 species up
  to $^{\rm 37}$Cl. The associated nuclear
  reaction rates for charged and neutron particle captures as
  well as beta decays have been kept up-to-date using the NETwork GENerator tool available at
  http://astropc0.ulb.ac.be (see \cite{Siess06} for details).  
  At low-temperature ($T < 8000$ K), the atomic
  and molecular opacities are given by
  \cite{AlexanderFergusson94}. Between $8000 \le T \le 5\times 10^8$
  K, we use the OPAL tables \citep{IglesiasRogers96}. 
The neutrino energy loss rates are computed according to \cite{Itoh96}
and take into account the effects of plasma, pair, bremsstrahlung,
recombination, and photo neutrino emission.

The equation of state
(EOS) is based on the principle of Helmholtz free energy minimization
and is described in detail in \cite{Siess00}. The atmosphere is
treated in the grey approximation and integrated up to an optical
depth $\tau \simeq 5\times 10^{-3}$.  Convection is modeled following
the mixing length formalism, and  a common parameter $\alpha_{\rm MLT} = 1.7378$ has been adopted for the three models
presented here, ensuring that the solar luminosity and radius are
reproduced with at most 0.1\% error (see Table 6).  We apply the
  Schwarzschild criterion for the convective instability. Convective
  regions are assumed to undergo instataneous mixing of chemicals
  and to rotate as rigid bodies. For the solar model, mass loss was
not taken into account.  The Maeder \& Zahn (1998) formalism is
  applied, and the resulting set of five non-linear differential
  equations is solved using a Newton-Raphson numerical scheme
  according to \cite{Henyey64} as described in \cite{PCTS06}.  The
  transport of nuclides by gravitational settling is also accounted
  for in all the rotating solar models. It is introduced using the
  approximation of \cite{Paquette86} for the microscopic diffusion
  coefficient and the expressions given by \cite{MM76} for the
  microscopic diffusion velocity.

\subsection{{\bf Assumptions introduced in the rotating models of the Sun}}\label{sec:hypo}

The Sun is the only star where we can perform quantitative comparison of
the impact of the above prescriptions on the sound speed and the rotation
profiles which can now be deduced from helioseismic data. So it is
interesting to discuss in details the role of the different terms and the
choice for the prescriptions. In the following section, we shall present
two types  of models:  one model with an extremely slow initial rotation which does not justify
external magnetic braking (models A or {\em slow}), and two  models with greater initial rotation and
magnetic braking at the surface (models B and C, or {\em intermediate} and {\em strong} rotation models).  The main
characteristics of these models are given in
Table~\ref{tab:models}. These  cases allow to discuss the
order of magnitude of the meridional circulation, the diffusion
coefficients and the gradient of the internal angular velocity profile,
together with their different roles in the building of the present solar
rotation profile.
    
  For these models we have
  adopted \cite{Mathis3} prescription for the horizontal component of the
  turbulent diffusivity D$_h$:
   \begin{equation} 
   D_h =\sqrt{ \left(\frac{\beta}{10} \right)\left(r^2\overline{\Omega}
  \right)\left[r | 2V_2 - \alpha U_2 |\right]}.
  \end{equation} 
  The vertical component of
  the turbulent diffusivity D$_v$ is from \cite{TalonZahn97} :
    \begin{equation} 
D_v = \frac{Ri_c}{N^2_T/(K_T~+~D_h)~+~N^2_\mu/D_h}\left(r \partial_r
  \overline{\Omega}\right)^2 
    \label{eq:Dv}
    \end{equation} 
   with $Ri_c = 1/6$ the critical Richardson
  number, $K_T$ the thermal diffusivity, $N_T$ and $N_\mu$ the chemical and
  thermal parts of the Brunt V\"ais\"al\"a frequency $N^2 =
  N_T^2+N_\mu^2$.
  
Referring back to \cite{TalonZahn97}, let us recall here that for the
  shear instability to develop, not only the Richardson criterion must be
  satisfied, but the sheared flow shall be turbulent. This is controlled by
  the Reynolds criterion. In our models, we used a critical Reynolds number
  $Re_c = 10 \nu_{mol}$. When $\nu_v < Re_c$, $D_v$ is replaced by
  $\nu_{mol}$ in the transport equations for angular momentum and nuclides.

 The choice of slow and mild rotators on the Zero Age Main Sequence
  (hereafter ZAMS) refers to the two
  identified populations of young stars as pin-pointed by \cite{Bouvier09}
  : the slow rotators for which the angular momentum evolves like
  $\Omega^3$ and the fast rotators for which the angular momentum varies
  like $\Omega$.
  
  Models A explore the extreme case of slow rotators on the
  Pre-Main Sequence (hereafter PMS) and on the ZAMS, with a surface equatorial velocity of
  about 2.2 km.s$^{-1}$ remaining almost constant from the ZAMS to the
  present age of the Sun. No braking is applied at the surface.  These
models do not actually reproduce observed stars, and can be
considered more as  academic cases used to show the respective impact of the
meridional circulation and shear turbulence on the shape of the angular
velocity profile even with such small velocities. They are also used to estimate interesting quantities relevant for the present observations.

Models B and C are mild rotators at their
  arrival on the ZAMS (approximately 20 km.s$^{-1}$ and 50 km.s$^{-1}$
  respectively). These models undergo magnetic braking at the arrival on
  the main sequence, similar to the solar-mass models by
  \cite{TalonCharbonnel2005}.\\ Rotation is included from the PMS in all
  our models. We assume solid body rotation as the initial state in the
  completely convective PMS star. When magnetic braking is applied at the
  arrival on the main sequence, we use the formalism developed by
  \cite{Kawaler88} \citep[see also][]{Bouvier97,Palacios03} in order to
  obtain the solar surface equatorial velocity when the model reaches 4.6
  Gyrs :
\begin{equation}
\left(\frac{dJ}{dt}\right) \, = \, -K \Omega^3\left(\frac{R}{R_\odot}\right)^{1/2}\left(\frac{M}{M_\odot}\right)^{-1/2}~~~~~(\Omega < \Omega_{sat})
\end{equation}
\begin{equation}
\left(\frac{dJ}{dt}\right) \, = \, -K \Omega \Omega_{sat}^2\left(\frac{R}{R_\odot}\right)^{1/2} \left(\frac{M}{M_\odot}\right)^{-1/2}~~~~~(\Omega \geq \Omega_{sat})
\end{equation}

 This formulation corresponds to a field geometry intermediate between a
dipolar and a radial field (Kawaler 1988), and it 
is widely used in the literature (Chaboyer et al. 1995; Krishnamurti et
al. 1997; Bouvier et al. 1997; Sills \& Pinsonneault 2000).
 The parameter $K$ in Kawaler's law is related to the magnitude of the
   magnetic field strength, and is adjusted accordingly with the adopted
   initial rotation velocity.  $\Omega_{sat}$  expresses the fact that the magnetic field generation
saturates beyond a certain evolutionary point. This saturation is actually
required in order to retain a sufficient amount of fast rotators in young
clusters, as originally suggested by Stauffer \& Hartmann (1987). Here we
adopt $\Omega_{sat} = 14 \Omega_\odot$ following Bouvier et al. (1997).
  Table~\ref{tab:models} presents the main characteristics of models A, B
  and C computed with STAREVOL and CESAM.

\begin{figure}
\begin{center}
\includegraphics[width=10cm]{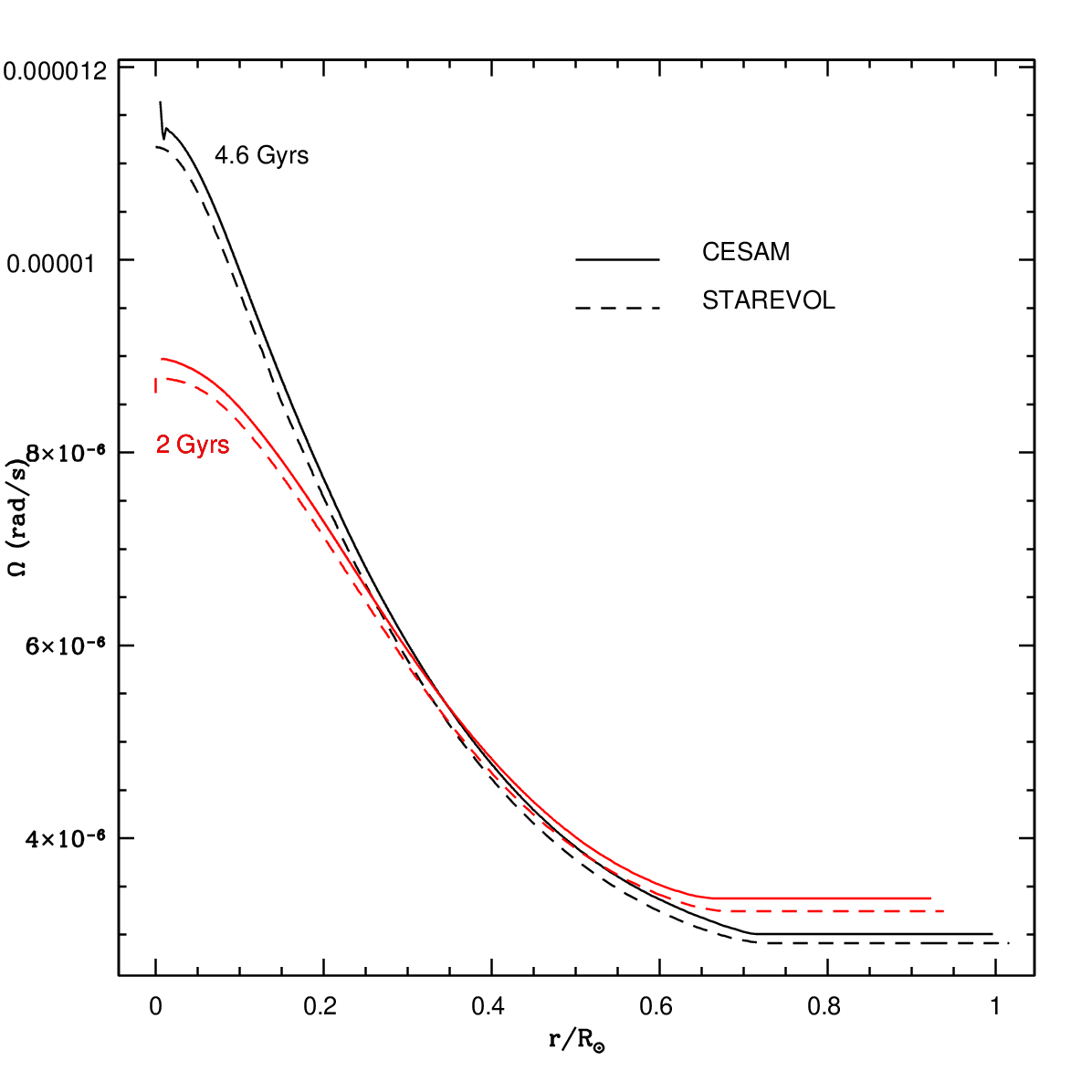}
\caption{Comparison of angular velocity profiles between the two codes CESAM (C) and STAREVOL (S) for  slow rotating models A$_C$ (solid
  lines) and A$_S$ (dashed lines) at 2 Gyrs and 4.6 Gyrs.}
\label{fig:comp_omega}
\end{center}
\end{figure}

\begin{figure}
\begin{center}
\includegraphics[width=8cm]{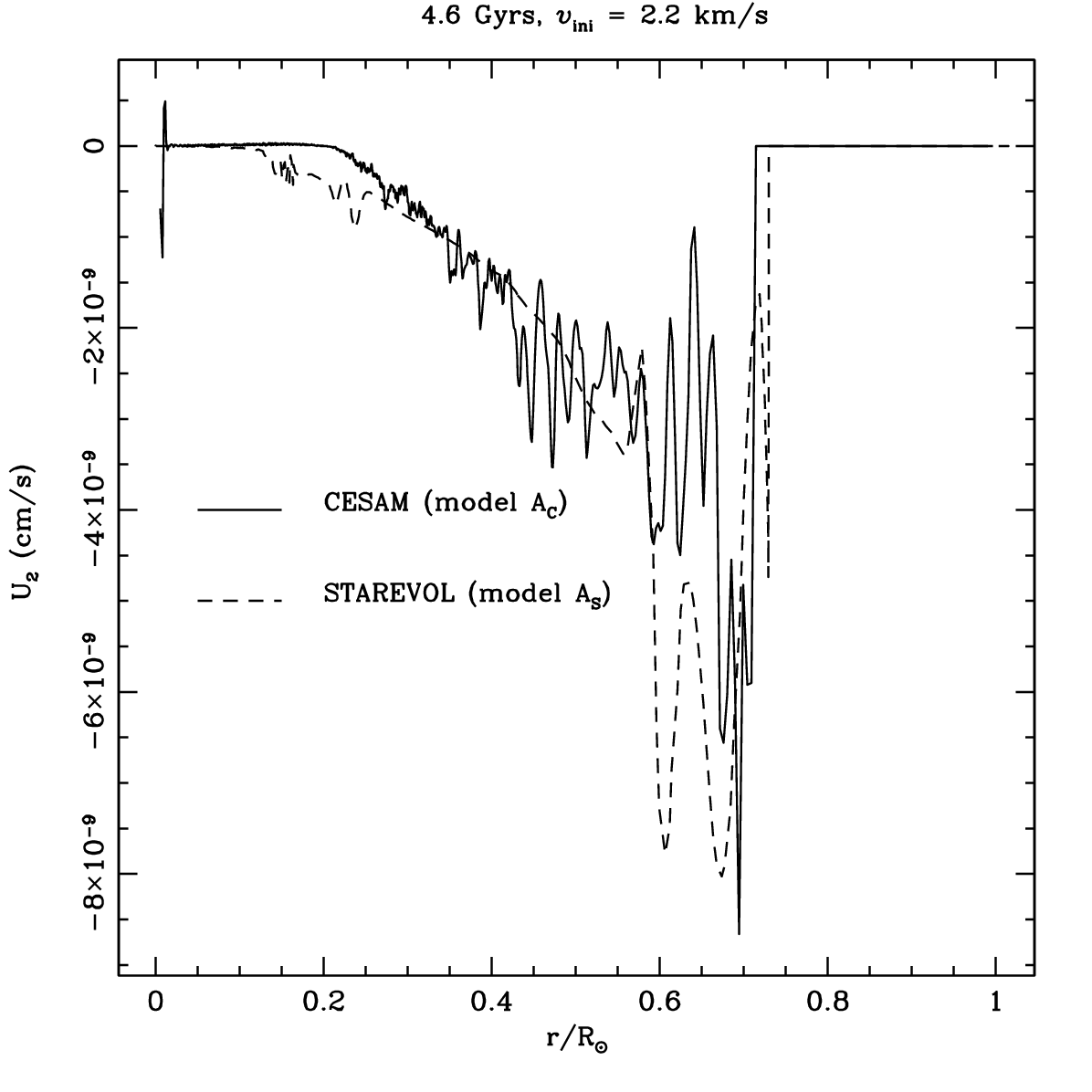}
\includegraphics[width=8cm]{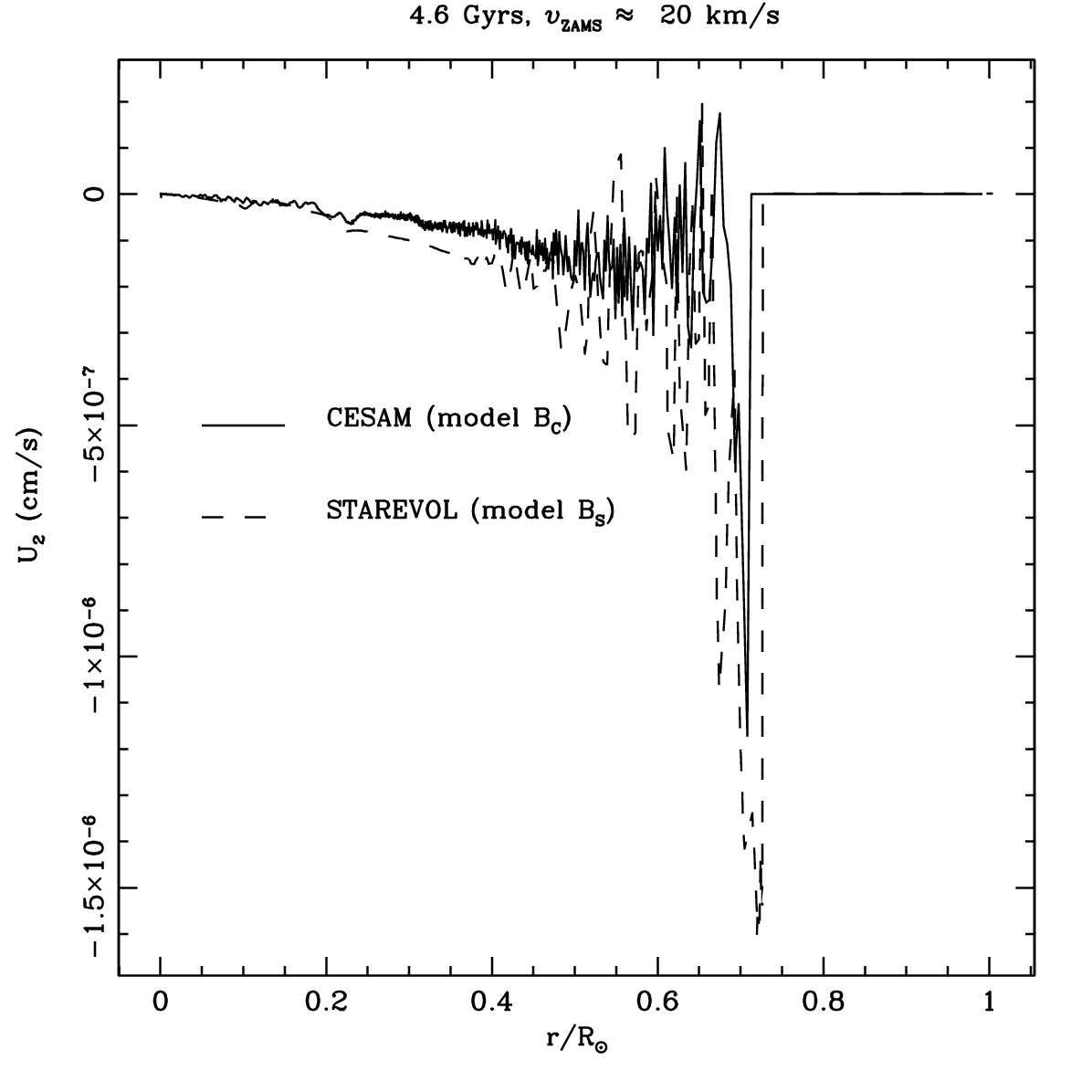}
\caption{Meridional circulation velocity profiles for solar models at 4.6 Gyrs computed with CESAM (solid lines) and STAREVOL (dashed lines) for very slow initial rotation (left) and mild initial rotation (right).}
\label{fig:comp_U2}
\end{center}
\end{figure}

\begin{table}
\caption{Characteristics of the calibrated models computed with CESAM for the slow initial rotation.}
\begin{tabular}
{|l|c|c|c|c|c|c|c|c|c|}
\hline 
 age	(Gyrs)    & R (R$_\odot$)	& L  ($L_\odot$) & E$_G$ (\%)  &
 E$_{pp}$   (\%) & E$_{CNO}$   (\%)&  $T_c$& $\rho_c$ & $\upsilon_{\rm c,
 eq}$ (km/s) & $\upsilon_{\rm s}$ (km/s)\\
\hline 
\hline 
0   		    &   17.2   &    73.       & 100    &   0     &     0 &  $4.83 \;10^5$   &   $2.1 \; 10^{-3}$  &     0.08     &    0.08    \\
0.00001          &     10.5  &  32.9      & 100  &  0  & 0  & $7.69 \;10^5$  & $8.5 \; 10^{-3}$ &      & 0.12  \\
0.0001      &      4.90       &        8.7            &        100        &    0        &   0    &     $1.6   \;10^6$    &    $7.8 \; 10^{-2}$  &        &   0.25      \\
0.001      &      2.20       &        1.8            &        100        &    0        &   0    &     $3.53   \;10^6$    &    $8.2\; 10^{-1}$  &        &   0.55      \\
0.015      &      1.12       &        0.46           &        100        &    0      &   0    &     $ 6.67   \;10^6$    &    $ 1.23 \; 10^{1}$  &     2.41   &   1.18      \\
0.020      &      1.02       &        0.63            &        94        &    6        &   0    &     $9.26   \;10^6$    &    $ 4.39 \; 10^{1}$  &        &   1.64      \\
0.030      &      0.96       &        0.89            &        33        &    65        &   0    &     $1.32   \;10^7$    &    $8.01 \; 10^{1}$  &        &   2.06     \\
0.039      &      0.88       &        0.71            &        33        &    92        &   8    &     $1.36   \;10^7$    &    $7.88 \; 10^{1}$  &        &   2.19     \\
0.050      &      0.87       &        0.70           &        0       &    94       &   6   &     $13.5   \;10^6$    &    $7.89 \; 10^{1}$  &     5.03   &   2.18      \\
0.147       &     0.88        &       0.72             &        0        &     98       &  2    &    $ 13.5    \;10^6$   &   $  8.26  \; 10^{1}$ &        &   2.18      \\  
1            &     0.9        &     0.77               &       0         &      100      &    0  &     $ 13.8    \;10^6$        &   91.2      &   5.6   &    2.177     \\
2             &      0.923       &        0.82            &       0         &      100      &   0   &        $ 14.2    \;10^6$      &   102.6     &  6.1    &     2.17    \\
3               &     0.948        &         0.88           &        0        &       100     &   0   &       $ 14.7    \;10^6$       &    111.7     &    6.3  &     2.16    \\
4.6            &       0.997      &              1.003     &       0         &      99      &   1   &     $ 15.65    \;10^6$         &    147.4     &   8.2   &    2.13     \\
\hline 
\end{tabular}
\label{tab:CESAMslow}
\end{table}

\section{THE ROTATING SOLAR MODELS and THEIR SENSITIVITY TO THE USED  PRESCRIPTIONS}

The computation of solar rotating models is
tricky due to the subtle changes induced by rotational mixing on the
internal structure of the Sun. The aforementioned formalism being
newly introduced in the CESAM code, we compare first in this section the
results obtained with STAREVOL and CESAM in order to achieve a mutual
validation of the results. The numerical approach to
solve the stellar structure equations in CESAM and STAREVOL is
significantly different and also is the numerical implementation of the
transport of angular momentum equation. We thus do not expect a perfect
match between the rotating solar models computed with each code, but orders
of magnitudes and shapes for the different profiles are expected to be
similar. 

\begin{figure}
\begin{center}
\includegraphics[width=8.cm]{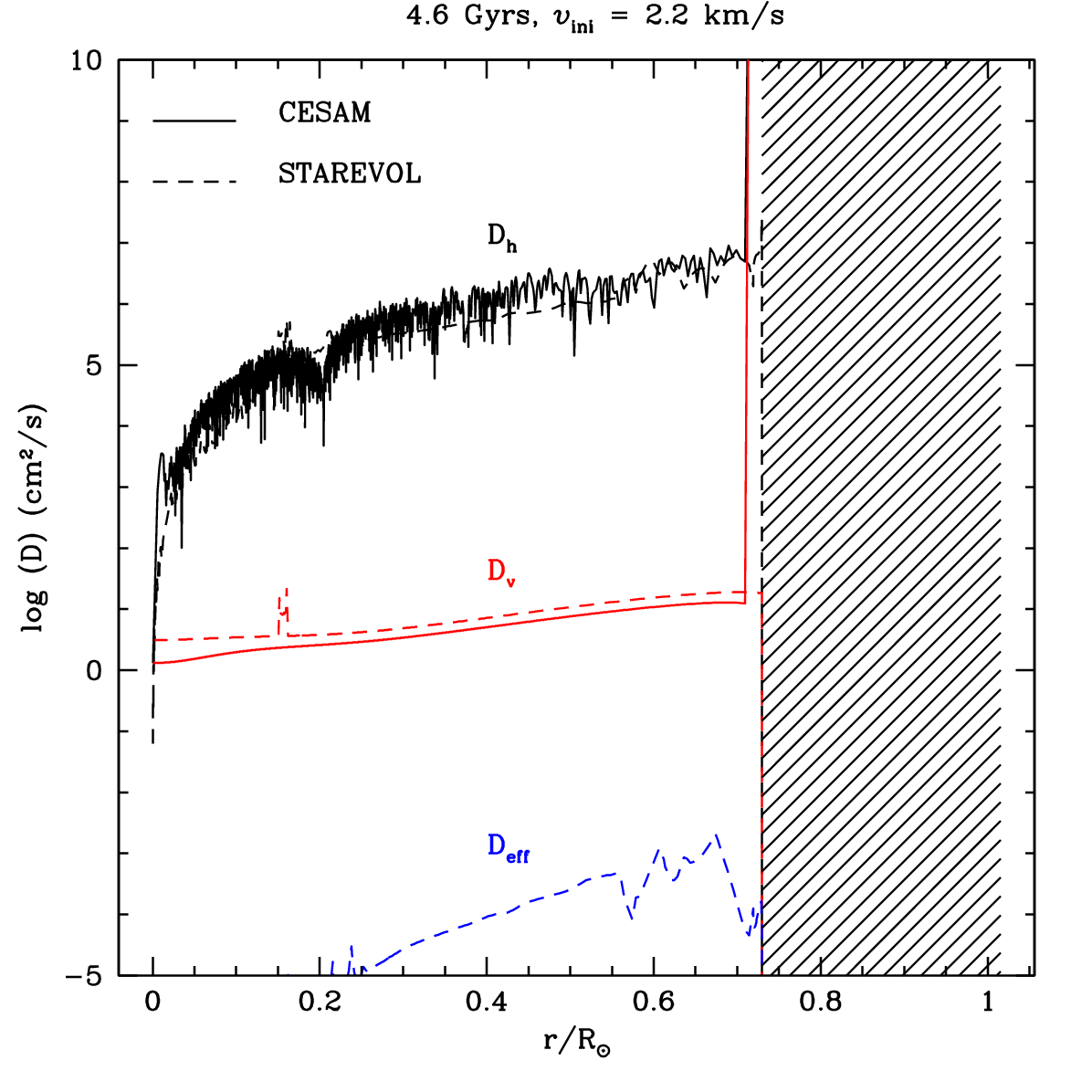}
\includegraphics[width=8.cm]{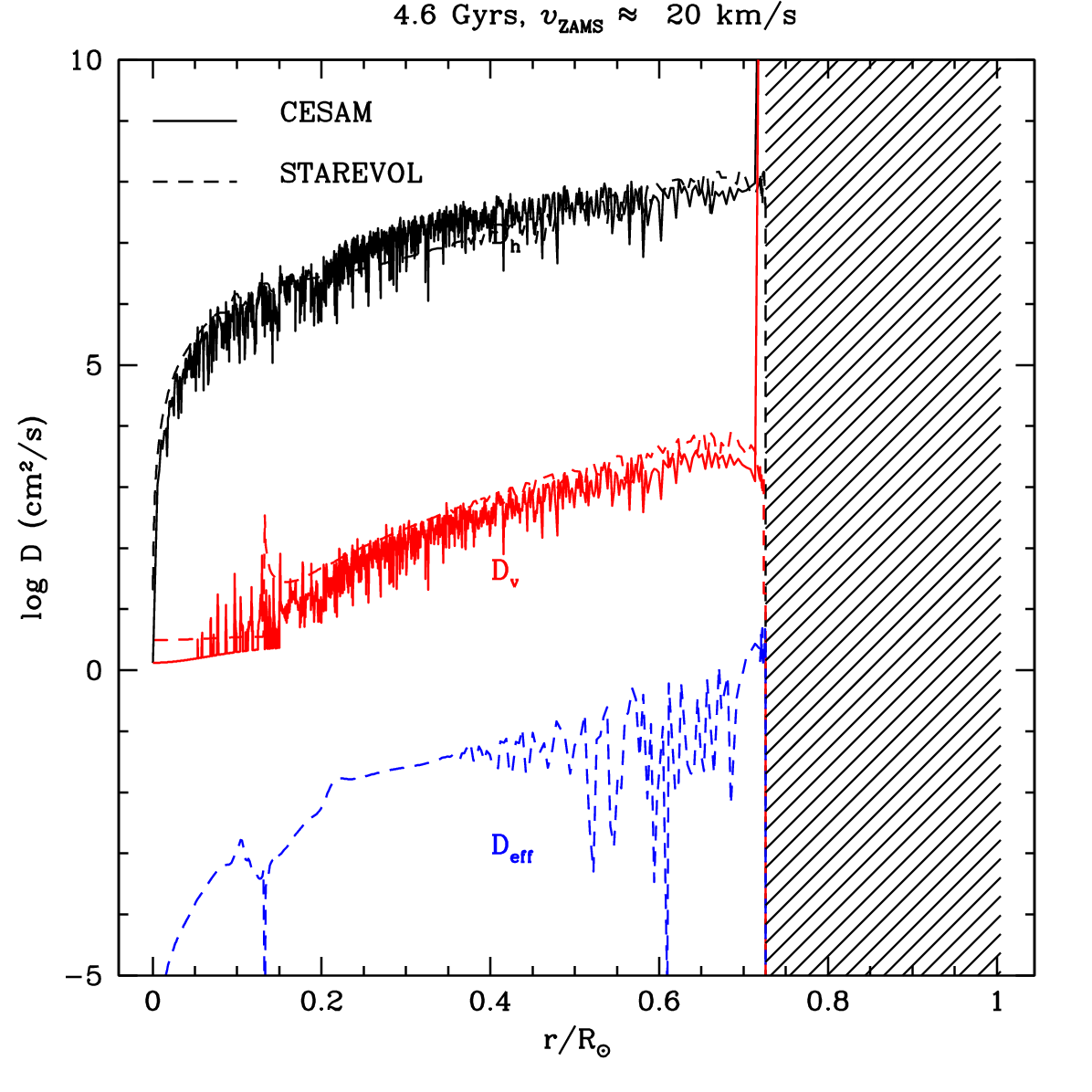}
\caption{Diffusion coefficient profiles for solar models at 4.6 Gyrs
  computed with CESAM and STAREVOL. The horizontal turbulent diffusion
  coefficient D$_h$, the vertical turbulent diffusion coefficient D$_v$ and
  the effective diffusion coefficient D$_{\rm eff}$ are presented for
  models A$_C$ (solid) and A$_S$ (dashed) on the left panel and for models
  B$_C$ (solid) and B$_S$ (dashed) on the right panel.}
\label{fig:comp_diff}
\end{center}
\end{figure}

\subsection{\bf An extreme case: the Sun, a slow rotator on the ZAMS}

We consider first the extreme case where the initial angular momentum
is small so that the model arrives on the ZAMS as a slow rotator
($\rm \upsilon_{ZAMS} \approx 2.2 km.s^{-1}$) without undergoing any magnetic braking.

Table~\ref{tab:CESAMslow} shows the evolution of several quantities
among them the surface velocity in the center and at the surface of
model A$_C$. With an initial rotation of about 0.1 km.s$^{-1}$, the
surface velocity is seen to increase to reach 2.19 km.s$^{-1}$ at the
arrival on the ZAMS (t = 390 Myrs), and then remains almost constant
up to the present age of the Sun. In the central regions, the increase
of the rotation velocity is maintained during the main sequence, so
that at the present age, $\upsilon_{\rm C}$ is about 4 times larger
than $\upsilon_{\rm S}$. This velocity gradient clearly appears in
Fig.~\ref{fig:comp_omega} and on the left panel of
Fig.~\ref{fig:evolomega}, where the evolution of the angular velocity
$\Omega$ is displayed for models A$_{\rm C}$ and A$_{\rm S}$. We may
note here the very good agreement between the profiles obtained with
CESAM and STAREVOL for solar models calibrated within 1\%, thus
validating the numerics of both codes. The building of the
$\Omega$-gradient is due on the PMS to the rapid contraction of the
central regions when radiation becomes the dominant energy transport
process in the core, as well as to the combined effects of meridional
circulation and  shear-induced turbulence. 

Despite the very slow rotation of these models, a small
temperature gradient is generated between the pole and the equator and a meridional
circulation develops in the radiative interior. The velocity of the
meridional currents, $U_2$ is of the order of
$\rm 10^{-9} cm.s^{-1}$ in absolute value at 4.6 Gyrs as can be seen in
Fig.~\ref{fig:comp_U2} left. This flow is extremely small compared to the
meridional circulation in the convective zone. The circulation consists in
a single counterclockwise loop, e.g. U$_2$ negative. The flow is directed downward, peaking near the
base of the convective zone, extracts angular momentum from the central
region to the base of the convective envelope. The profiles obtained with both codes are of
similar shape and amplitude. They present irregularities mainly due to the
mean molecular weight variations $\Lambda = \frac{\tilde{\mu}}{\mu}$ that
introduce non-linearities in the system because some noise is coming from
the mean molecular weight gradient. Let us here recall that the transport
of angular momentum introduced in the codes is treated in a self-consistent
but simplified manner. Such models give hints, order of magnitudes and
global shapes, but should not be expected to deliver exact profiles.

\begin{figure}
\begin{center}
\includegraphics[width=12cm]{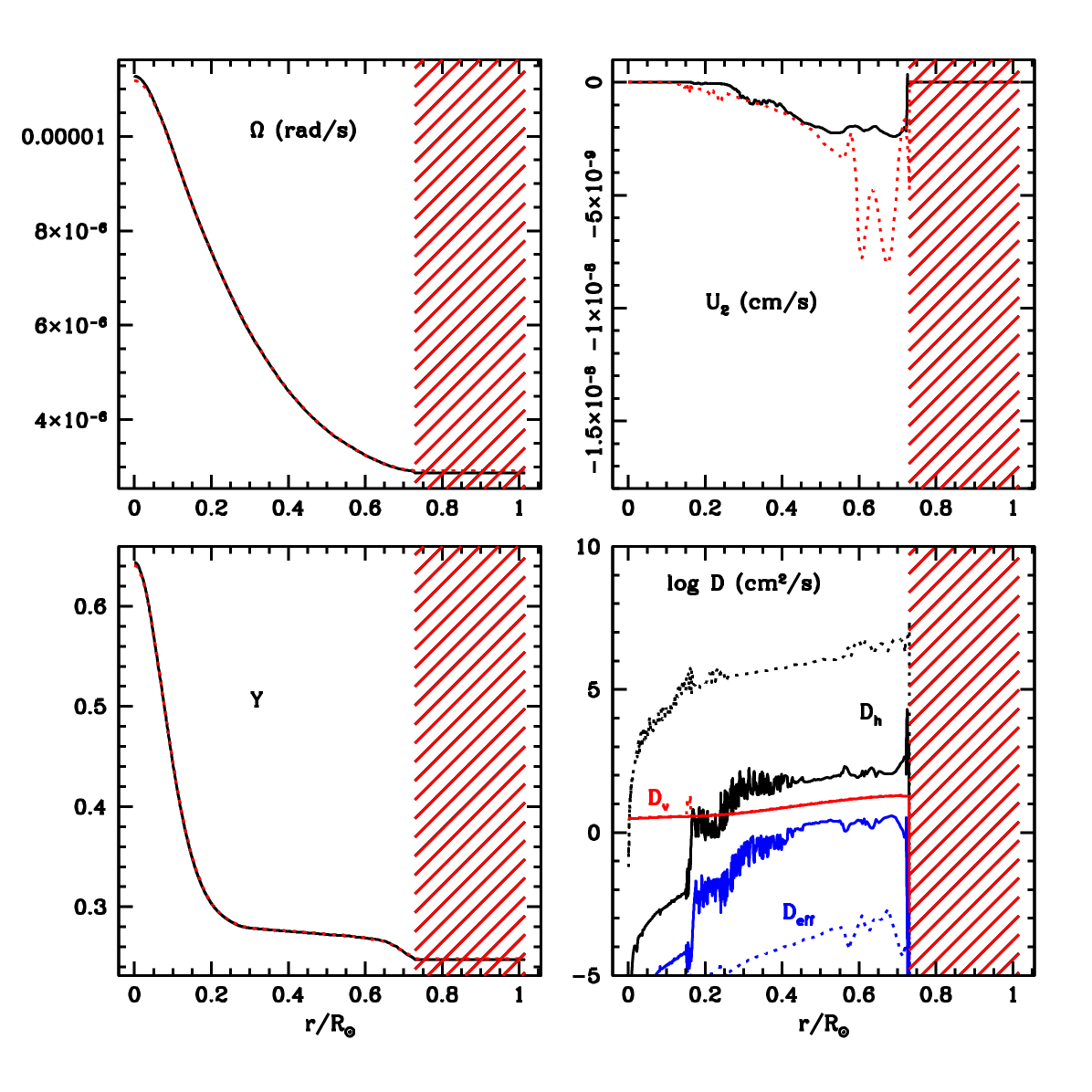}
\caption{Profiles of the angular velocity $\Omega$, the meridional
  circulation velocity U$_2$, the helium mass fraction $Y$ and the
  diffusion coefficients for the solar rotating models A$_S$ (dotted lines) and A'$_S$ (solid
  lines) at 4.6 Gyr. The hatched areas indicate the convective envelope.}
\label{fig:ZahnvsMPZ}
\end{center}
\end{figure}

The diffusion coefficient profiles are also similar in the two codes (see
Fig.~\ref{fig:comp_diff} left). The effective diffusion coefficient
D$_{eff}$ is proportional to $(U_2)^2/D_h$ and consequently is very small (about 10$^{-5}$ to 10$^{-3}$ $\rm cm^2.s^{-1}$)
due to the slow meridional circulation velocity and the large horizontal
component of the turbulent diffusivity ensured by the use of \cite{Mathis3}
prescription. The Reynolds number of the sheared flow is lower than the
critical Reynolds number Re$_c \propto 10 \nu_{mol}$ in the bulk of the
radiative zone. The flow is not turbulent, so the vertical component of the
turbulent diffusivity $D_v$ is essentially set equal to the local molecular
viscosity (see \S~\ref{sec:hypo}). The effective diffusion coefficient is
actually much smaller than the microscopic diffusivity \citep[see e.g.][]{Brun2}, which is of the
order of 10 $\rm cm^2.s^{-1}$ in solar models. Consequently the microscopic diffusion
is the most efficient process to transport chemical species below the
convective envelope in the models with slow rotation. This shows up in the helium mass fraction profile presented in quadrant III
of Fig.~\ref{fig:ZahnvsMPZ}. There, the step in the helium profile at the
base of the convective zone reveals the gravitational settling
of helium in this region, and this, regardless of the existence of
meridional flows and shear. If the young
Sun was a slow rotator, the chemical stratification probed by helioseismic
data in the region below the convective envelope would be the same as if
the Sun had not rotated (provided that no other dynamical process than
rotation is included).\\ Of course, one
cannot ensure that the diffusivity coefficients are properly estimated, so
it is interesting to see the impact of a different prescription for them.
Using the STAREVOL code, we compute model A'$_S$ using the prescription of \cite{Zahn92} for D$_h$:
\begin{equation}
D_h = r
\left[2 V_2 - \alpha U_2\right]
\end{equation}

 Figure~\ref{fig:ZahnvsMPZ} shows the profiles of the angular velocity
$\Omega$, the meridional circulation velocity $U_2$, the helium mass
fraction $Y$ and the diffusion coefficients D$_h$, D$_v$ and D$_{\rm eff}$ for
models A$_S$ and A'$_S$. As expected, the
horizontal turbulent diffusion coefficient D$_h$ is three orders of
magnitudes smaller when using Zahn's 1992 prescription \citep[see
also][]{Mathis3}. This translates into an increase by the same amount of
the effective diffusivity D$_{\rm eff}$. However, D$_{\rm eff}$ remains smaller
than D$_v$ and $\nu_{\rm mic}$ in model A'$_S$, so that the helium and angular
velocity profiles reamin unchanged compared to model A$_S$. From this comparison, we
may conclude that in the case of a Sun that has always been rotating
slowly, the choice of the prescription for the horizontal component of the
shear-induced turbulent diffusivity does not affect significantly the
structure, rotation and initial chemical stratification.

\begin{figure}
\begin{center}
\includegraphics[width=12cm]{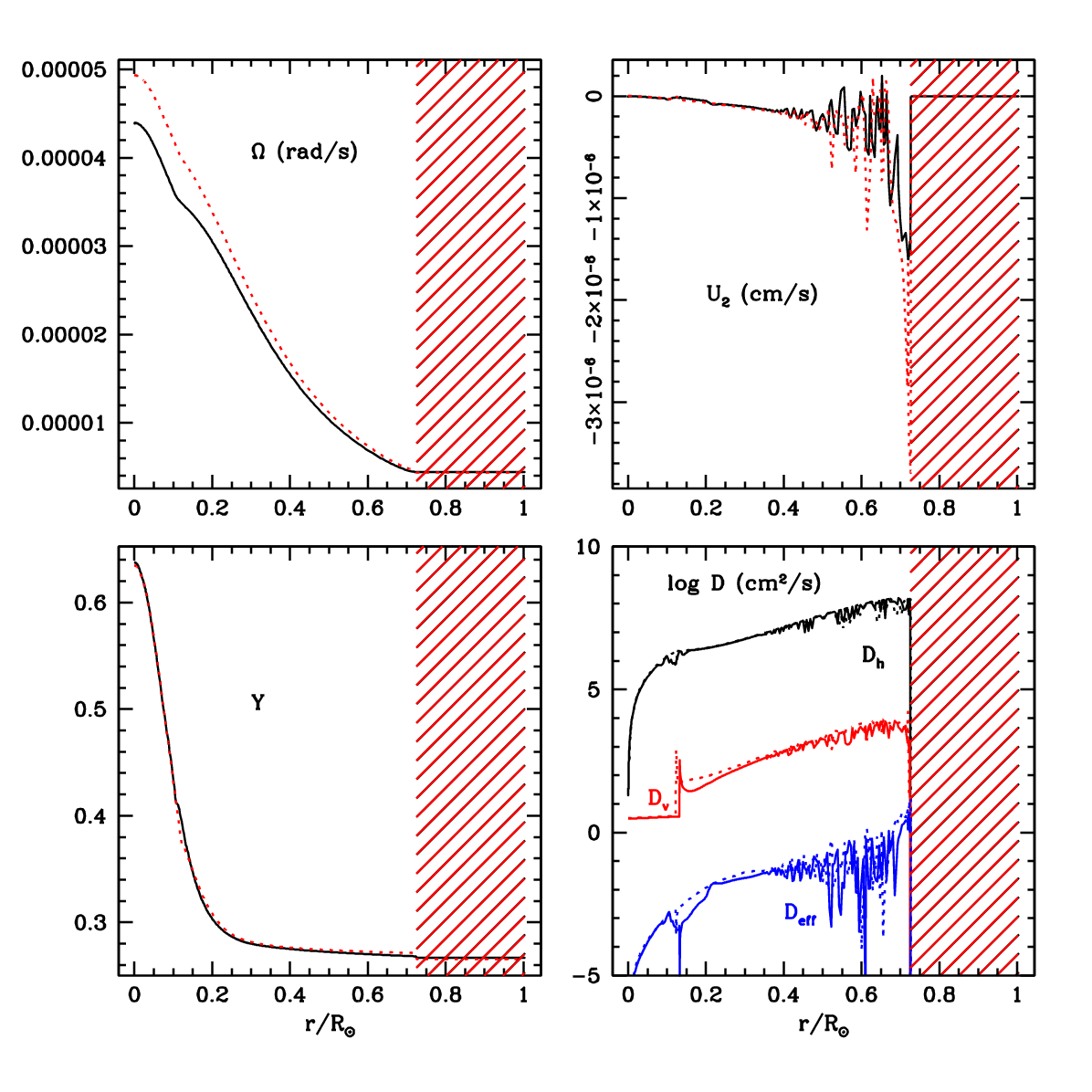}
\caption{Profiles of the angular velocity $\Omega$, the meridional
  circulation velocity U$_2$, the helium mass fraction $Y$ and the
  diffusion coefficients for the solar rotating models B$_S$ (solid lines) and C$_S$ (dotted
  lines) at 4.6 Gyr. The hatched areas indicate the convective envelope.}
\label{fig:20vs50}
\end{center}
\end{figure}

\subsection{{\bf The Sun as a mild rotator on the ZAMS}}

 We have also computed two models with a higher initial angular
 momentum,  corresponding to a surface velocity of about 20 km.s$^{-1}$ and 50
 km.s$^{-1}$ at the arrival on the ZAMS. These models undergo a strong
 angular momentum extraction on the early main sequence to reach a
 surface velocity of about 2 km.s$^{-1}$ at the age of the present
 Sun. This phenomenon associated to the idea of a magnetic braking due
 to the decoupling from the disk environment is modelled using the
 Kawaler formula as detailed in the previous
 section.\\ Fig.~\ref{fig:20vs50} compares the results of the two
 models B$_S$ and C$_S$ at 4.6
 Gyrs. The diffusion coefficients D$_v$ and D$_{\rm eff}$ entering the
 diffusion equation for the chemical species are similar in models
 B$_S$ and C$_S$, so they lead to a same profile for the helium mass
 fraction. The angular velocity is slightly larger in the central
 regions in model C$_S$. This model is submitted to a more efficient
 braking in order to reach a surface velocity of about 2-2.2
 km.s$^{-1}$ at the age of the Sun. In fact, when magnetic braking is
 initiated, the meridional circulation velocity $U_2$ is five times
 smaller in model C$_S$ than that found in model B$_S$ in absolute
 value, but it is positive (negative in model B$_S$) generating an
 efficient {\em inward} transport of angular momentum and leading to a
 faster and larger increase of the central velocity than in model
 B$_S$. The behaviour found for model C$_S$ is in all points similar
 to that reported by \cite{Eggenberger05}, and also resembles that
 described by \cite{MeynetMaeder00} for more massive and fast rotating
 stars. The evolution further on the main sequence is similar for
 models B$_S$ and C$_S$, with $U_2$ rapidly becoming negative in the later (see
 figure top right of Fig.~\ref{fig:20vs50}). The profiles displayed in
 Fig.~\ref{fig:20vs50} show no significant difference for the
 diffusivities, meridionla circulation and helium profile at 4.6 Gyrs, so
 that in the following we will focus on models B$_S$ and model B$_C$.

The meridional velocity profiles at 4.6 Gyrs  for models B are shown on the right panel
of Fig.~\ref{fig:comp_U2}. The velocity of the meridional currents is 3
orders of magnitude larger compared to models A$_S$ and A$_C$ previously
described. At the ZAMS, the meridional circulation is maximal below the
convective envelope as a result of the strong magnetic torque applied at
the surface. $U_2$ is then 2 orders of magnitude larger than it will be
later on the main sequence, when the efficient spin down of the surface
layers is accompanied by a global weakening of the meridional
circulation. The form of the meridional circulation velocity profile $U_2$
is maintained over the main sequence evolution, and
mainly  consists in one counterclockwise cell peaking below the convective envelope
and extracting angular momentum from the central regions, this time more efficiently than in
the slowly rotating solar models. Similar to what was obtained by
\cite{Decressin09} for their 1.5 M$_\odot$ model, the meridional
circulation is mainly driven by the local losses of angular momentum due to
the magnetic braking. Meridional circulation and shear turbulence do not
ensure an efficient transport of angular momentum.
One notes that $U_{2,max}$ is larger in model B$_S$ by 25$\%$ compared to model B$_C$.

\begin{figure}
\begin{center}
\includegraphics[width=9cm]{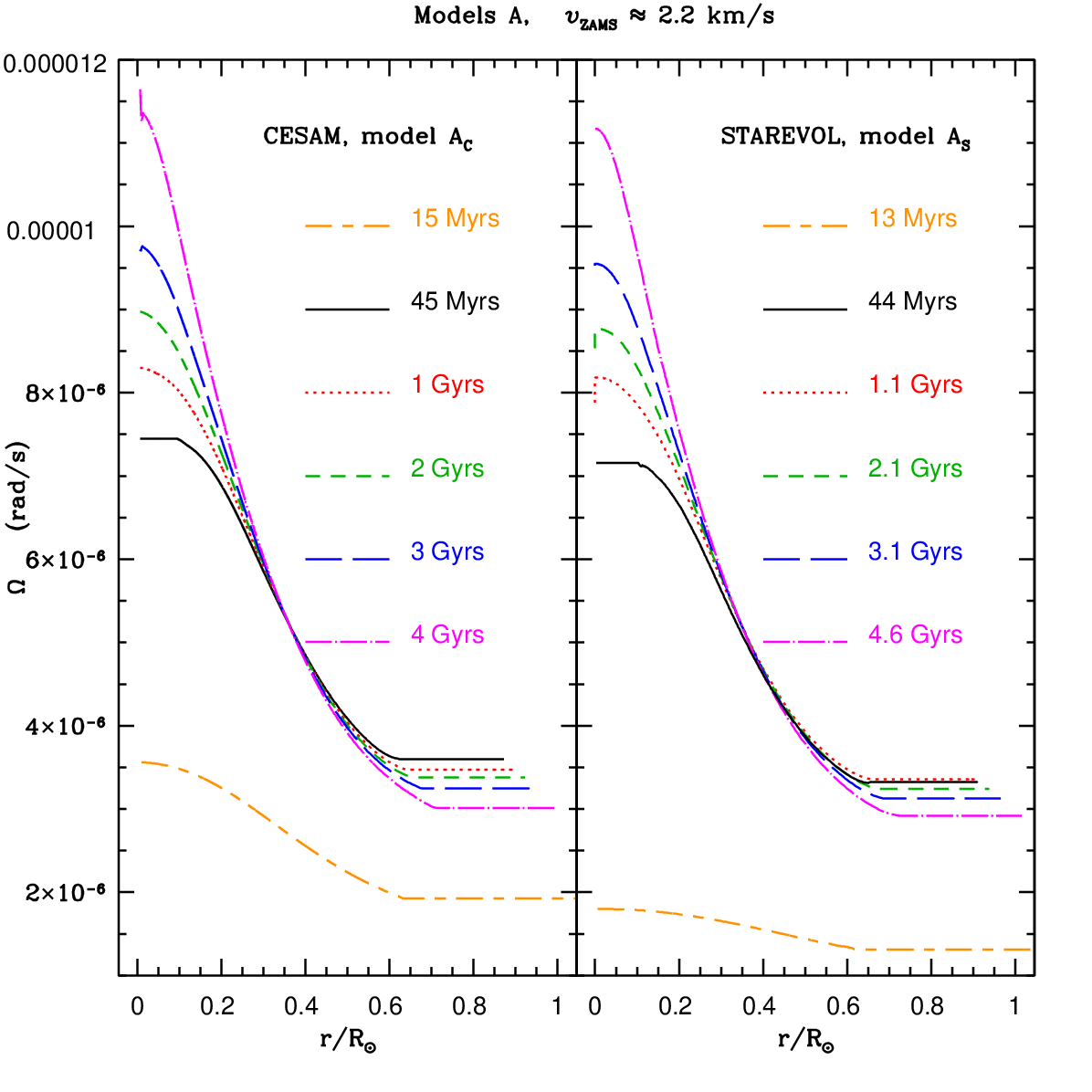}
\includegraphics[width=9cm]{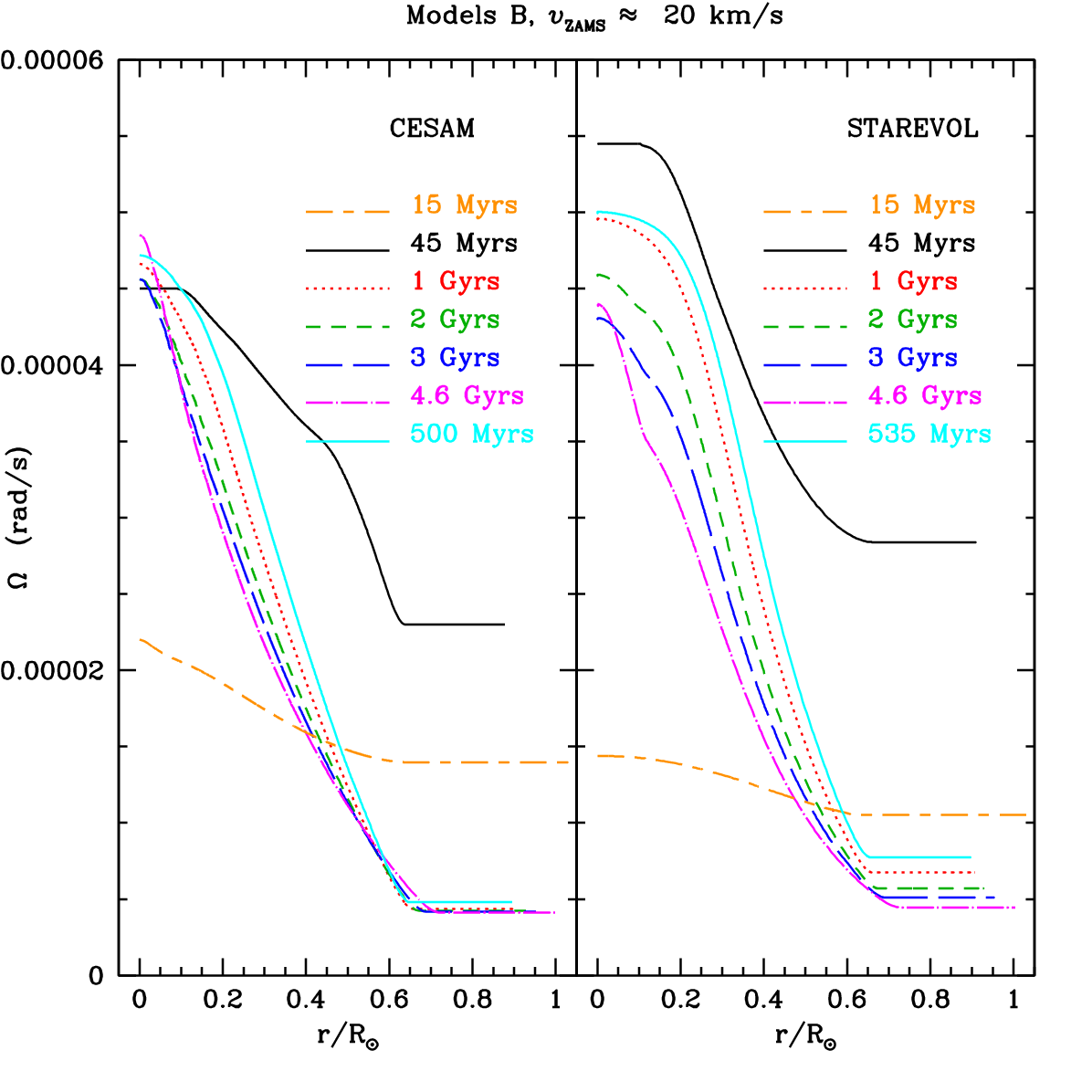}
\caption{Evolution of the angular velocity profiles at six different
  ages as indicated on the plots, for models A$_C$ and A$_S$ (left
  panels) and B$_C$ and B$_S$ (right panels). The age of 45 Myr
  corresponds to the arrival on the main sequence.}
\label{fig:evolomega}
\end{center}
\end{figure}

The gradient of angular velocity is much larger in these models compared to
models A$_C$ and $A_S$, and $\Omega$ increases by more than one order of
magnitude from the surface to the solar core. We will discuss in more
details the implications of such a steep profile when confronted to
helioseismic data. The match between the two codes shown in
Fig.~\ref{fig:evolomega} is not as good as that found for slowly rotating
models. This is due to the difference in the numerical treatment used in
both codes and the way the braking is introduced. The rotation profile is even
steeper for model computed with CESAM, in which braking occurs very
efficiently on the ZAMS so that the $\Omega$ profile does not evolve at all
in the last 3.6 Gyrs. In model B$_S$, angular momentum is extracted from
the central parts as the star evolves and both the central and surface
values of the angular velocity decrease with increasing time so that the
gradient is less steep at 4.6 Gyrs than it is at 1 Gyr.

Diffusion coefficients are displayed in the right panel of
Fig.~\ref{fig:comp_diff} at 4.6 Gyrs. They have similar shapes and
amplitude, which implies that the chemical stratification should also be
similar in models B$_S$ and B$_C$. Although U$_2$ is larger by 3 orders of
magnitude compared to models A, the horizontal component of the turbulent
diffusivity is also much larger leading to D$_{eff}$ that is again smaller
than 1 cm.s$^{-1}$, which remains much smaller than both the local
molecular diffusivity and the microscopic diffusivity. Meridional
circulation is thus not contributing significantly to the transport of
chemical species. On the other hand, in these mild rotating models, the
Reynolds number associated with the sheared flow is larger than the
critical Reynolds number $Re_c$, and the flow eventually becomes
turbulent. The vertical turbulent diffusivity reaches values of the order
of 100-1000 cm.s$^{-2}$ below the convective envelope, which is at least 10
times larger than the microscopic diffusivity in this region. As a
consequence, the abundance profiles are flattened in both
 models B$_S$ and B$_C$ (see e.g. see solid line in quadrant III of Fig.~\ref{fig:20vs50} for
model B$_S$), and differ thus significantly from those derived for standard
solar models. If the young Sun was a fast rotator and experienced magnetic
braking during the early main sequence without any inhibition of the action
of the shear-induced turbulent mixing, the chemical stratification of the
present Sun below the convective zone might bear the signature of this past
rotational history.

\begin{table}
\caption{Characteristics of the calibrated models computed with CESAM for an
  initial rotation of 20 km/s. }

\begin{tabular}
{|l|c|c|c|c|c|c|c|c|c|}
\hline 
 age	(Gyrs)    & R (R$_\odot$)	& L  ($L_\odot$) & E$_G$ (\%)  &
 E$_{pp}$   (\%) & E$_{CNO}$   (\%)&  $T_c$& $\rho_c$ & $\upsilon_{\rm
 c,eq}$ (km/s) & $\upsilon_{\rm s}$ (km/s)\\
\hline 
\hline 
0               &        17. 36             &            72.4               &        100          &      0         &      0    &                     0.48      $\;10^5$               &    2.05  $\;10^{-3}$          &         0.47          &          0.47          \\ 
0.00001  &   10.62    &   32.5    &     100   &   0      &     0     &     7.58 $\;10^5$     &   8.26  $\;10^{-3}$  &           &   0.74    \\
0.0001   &    4.95    &    8.6    &   100    &    0    &    0     &  1.58  $\;10^6$     &    7.55 $\;10^{-2}$  &      &    1.51    \\
0.001  &   2.28    &    1.92   &    100   &   0   &  0   & 3.38  $\;10^6$     &  0.73   &      &   3.18  \\
0.005  & 1.36  &  0. 63     &   100    &   0   &   0    & 5.32   $\;10^6$  &   4.1    &       &     5.51  \\
0.011   &   1.12   &  0. 46   &  100   &   0   &   0    &    6.62   $\;10^6$ &  12.15    &    &   7.99  \\
0.026    &    1.024   &  0.897    &   69    &   31    &    0    &   11.9   $\;10^6$  &  75.53   &     &    16.78  \\
0.039  &   0.88   &   0.71  &   0   &    93    &   7    &    13.58 $\;10^6$   &  79.18  &   &  20.25  \\
0.045       &     0.87       &       0.69             &          0         &    93      &   7    &    $ 13.56     \;10^6$  &  78.68    &      26.7        &       19.6      \\
0.147       &     0.88        &       0.72             &        0        &     98       &  2    &    $ 13.5    \;10^6$   &    82.6  &    30.8    &      5.   \\  
1            &     0.9        &     0.76               &       0         &      100      &    0  &     $ 13.8    \;10^6$        &   91.      &      &    2.19     \\
2             &      0.921       &        0.82            &       0         &      100      &   0   &        $ 14.2    \;10^6$      &   102.4     &    38.8  &     2.08    \\
3               &     0.945        &         0.88           &        0        &       100     &   0   &       $ 14.7    \;10^6$       &    116.4     &   37.3   &     2.086    \\
4.6            &       0.992      &              1.000     &       0         &      99      &   1   &     $ 15.61    \;10^6$         &    146.6     &   39.   &    2.09     \\
\hline 
\end{tabular}
\label{tab:CESAM20kms}
\end{table}

\section{CONFRONTATION BETWEEN THEORETICAL AND OBSERVATIONAL  PROFILES}

\subsection{The solar observed rotation profile}

After more than 10 years of observations with GOLF and MDI instruments
located onboard SoHO, and with the facilities of the GONG groundbased network,
we have derived very important constraints on the radiative zone rotation
profile.  The determination of the splittings of a
large number of acoustic modes has definitely
established that the rotation profile in the part of the radiative zone which is not
influenced by the nuclear reaction rates, e.g. the region
between 0.25 to 0.68 R$ _\odot$, is really flat with invisible
latitudinal differential variation ~\citep{Couvidat,Eff-Darwich2008},
in great contrast with the latitudinal differential profile of the
convective zone. In the central region, below 0.25 R$ _\odot$, only
gravity modes may inform on the rotation profile. Some individual
candidate modes at high frequency have been observed by the GOLF
instrument ~\citep{Turck04b,Mathur,Garcia08}, and two solutions have
been extracted from these first data depending on the interpretation
of the pattern attributed to an $\ell$ = 2, n = 2 mode: a slightly
reduced rotation rate or an increase by about a factor 3  in the
central region \citep{Mathur2}.  Nevertheless, only the solution of a
rapid rotating core is compatible with the other kind of detection
using the asymptotic behaviour at low frequency which is detected with
a very high probability~\citep{Garcia2007,Garcia08}. It could presume a rather
complex solar core rotation larger than the rest of the radiative
zone, with a different axis, and may be with some manifestation of the
radiative zone magnetic field. The other solution cannot be totally excluded
but is not the most probable. Again, the core profile needs to
be confirmed with extensive data from SoHO and with an improved
instrument like the GOLF-NG concept ~\citep{Turck2006,Turck2008} which
will measure velocity displacements at 8 heights in the solar
atmosphere. 
Considering that the complete solar rotation profile might
be accessible to the observation, it is important to predict it properly
down to the central region.

\subsection{The theoretical solar rotation profile}                      

This paper is a new step toward a detailed understanding on how the
different dynamical processes can influence the present rotation of
the Sun. In this study we have followed three different cases because
we can only have some indirect information on the solar internal
rotation profile at the end of its contraction phase. We have already
seen in the previous section that the initial angular momentum content
and the rotation history, in particular the fact of undergoing
magnetic braking on the ZAMS, shape the angular velocity profile,
determining the absolute value as well as the gradient.  In this
section, we wish to focus on the evolution of the angular velocity
profile. We present hereafter a detailed analysis of the construction
of the predicted internal profile at the age of the Sun and confront
these predictions to helioseismology.

Compared to previous studies by \cite{Eggenberger05} and \cite{TalonCharbonnel2005}
where rotation was only included from the ZAMS, with an initial flat
profile, we have decided to follow rotation from the PMS when the star is
actually completely convective. We also assume solid-body rotation as the
initial state, but by the time the models reach the ZAMS, the convective
region has retreated to the surface and the $\Omega$-profile is not flat
anymore.

Tables~\ref{tab:CESAMslow} and ~\ref{tab:CESAM20kms} list the
evolution of the radius, the luminosity, the central and the
superficial velocity for models A$_C$ and B$_C$ and some other
indicators of the contraction phase and of the involved nuclear
reactions. The evolution of the angular velocity profile for these
models is also presented in the left panels of
Fig.~\ref{fig:evolomega}.

For model A$_C$, where the initial rotation rate is of about 0.08 km.s$^{-1}$, the
contraction of the star  during the PMS rapidly
generates a radial gradient in the $\Omega$ profile.
The very slow meridional currents and the small amount of shear generated
in such slowly rotating models are not very efficient to transport the
angular momentum. The differential rotation established during the early
evolution by the contraction of the inner regions is maintained and
slightly amplified during the main sequence  by the advection term. The predicted overall
contrast between core and surface velocities is of about a factor of 4,
which is comparable to that obtained by \cite{Eggenberger05} in their
slowly rotating model.

\begin{figure}
\begin{center}
\includegraphics[width=9.cm]{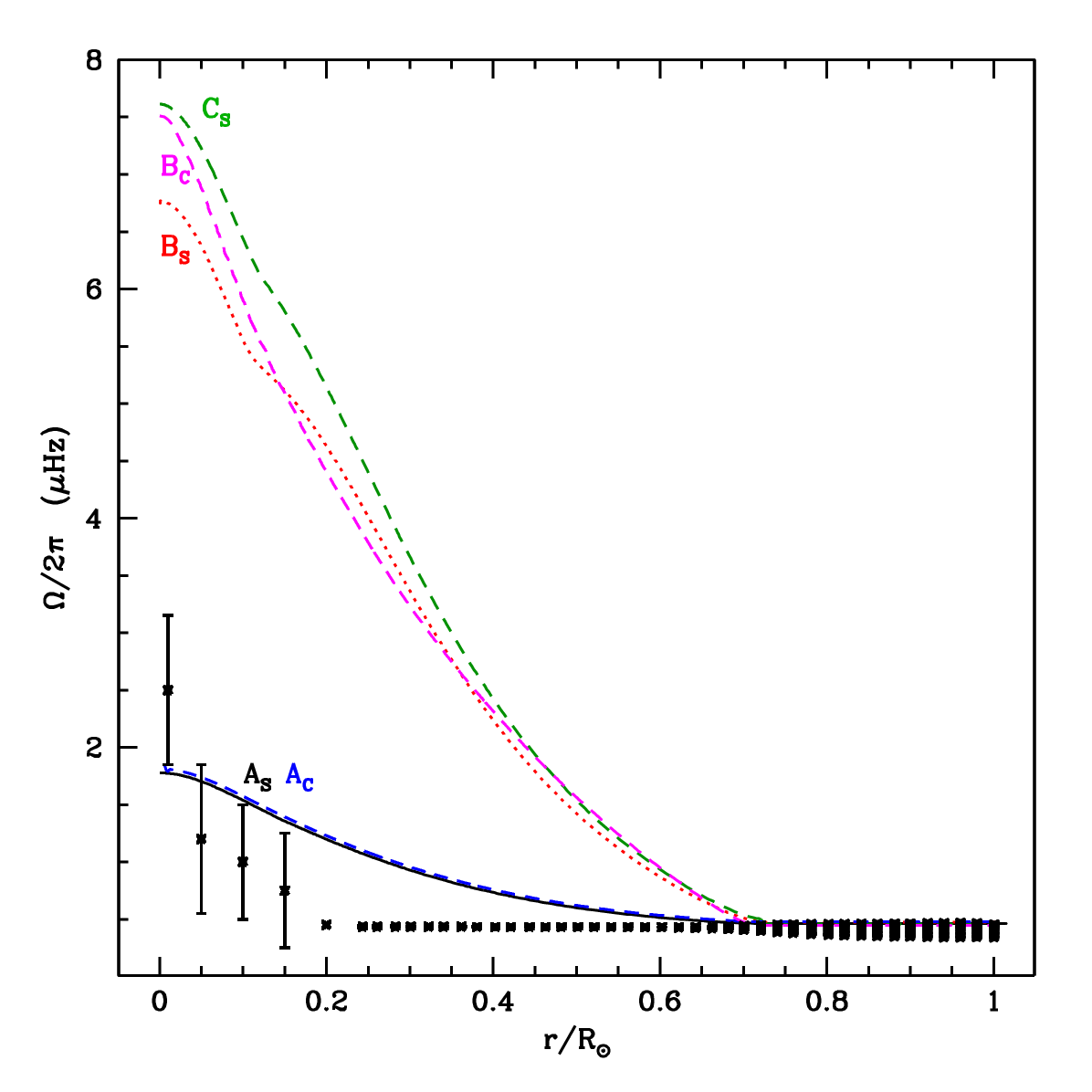}
\caption{Comparison between the solar internal profile predicted by
    the different models and that deduced from helioseismology. For models B$_S$ and 
C$_S$, the actually ploted value is $\Omega/2\pi$-0.2325 rather than 
$\Omega/2\pi$ in order for the surface value they reach to match that 
reached by model B$_C$. The data points down to r/R$_\odot$ = 0.2 are deduced  from the acoustic mode splittings determined
by the observations of GOLF, MDI and GONG instruments (from Eff-Darwich,
2008). The data points in the core correspond to the supposed core rotation extracted from the potentially
observed gravity modes. These values have still large error bars and need to
    be confirmed  (inspired by Turck-Chi\`eze et al.,2004b, Garcia et al. 2007 and
Mathur et al. 2008). }

\vspace{-0.5cm} 
\label{fig:sismo}
\end{center}
\end{figure}

Models B and C are similar to those presented in the three main previous
studies of the effect of rotation on the solar evolution by
\cite{Chaboyer,Eggenberger05,CharbonnelTalon05}. The evolution of the
angular velocity profile is in agreement with that obtained in the later
study, where STAREVOL models were also used. Comparing the evolution
displayed in Fig.~\ref{fig:evolomega} with that of Fig.~1 and Fig.~2 of
\cite{CharbonnelTalon05} and \cite{Eggenberger05} respectively, one can
measure the impact of the numerical approach. In all these cases, if the
general forms and amplitudes of the angular velocity profile at the age of
the Sun show encouraging similarities (see also Fig.~\ref{fig:sismo}), the
in between evolution can be quite different.
Models B$_S$ and B$_C$ undergo a very strong braking at the arrival on the ZAMS
leading to a quick spin down of the convective envelope and a
contrast between the surface and central angular velocities of about a factor 15 for model B$_C$ already at 1 Gyr, and of
about 7 in model B$_S$ at the same age. In the later model, the contrast
increases during the main sequence to reach a factor of 10 between the core
and the surface at 4.6 Gyrs. In both models, the $\Omega$-gradient steepens in the
bulk of the radiative zone as the star evolves. The profiles are flater in model
B$_S$, yet not as flat as those presented in Fig.~2 of \cite{Eggenberger05}.

 A detailed
  analysis of the relative importance of meridional circulation versus
  shear-turbulence in our models A$_S$, B$_S$ and C$_S$ using the
  tools presented in \cite{Decressin09}, demonstrates that the
  behaviour of these solar models is in all points similar to that
  obtained for the 1.5 M$_\odot$ model shown in that paper. The
  angular momentum transport is ensured by meridional currents. In the
  case of models B and C, these currents, although slow, are primarily
  generated by the action of the applied torques resulting from the
  action of magnetic braking. In the case of the slow model A, no
  torque is applied and the angular momentum flux is smaller yet also
  dominated by the meridional circulation. The flux of angular
  momentum due to shear turbulence is more than 5 orders of magnitude
  smaller than that attributed to the meridional circulation in all
  the 3 cases.
             
\begin{table}
\caption{Comparison of the calibrated solar models computed with the CESAM  code and those of the
   STAREVOL code. X$_{i}$ and Y$_{i}$ are respectively the initial hydrogen and helium mass fractions. X$_{c}$, Y$_{c}$, T$_{c}$ and $ \rho_{c}$ are respectively  the present central hydrogen, helium, temperature and density. Y$_{s}$ is the present superficial helium mass fraction, $\alpha$ is the mixing length parameter, r$_{BZC}$ the position of the base of the convective zone, (Z/X)$_s$ the mass fraction of the heavy element in considering the GN composition.}
   \vspace{0.5cm}
\begin{tabular}{|l|c|c|c|c|c|c|c|}
\hline
& CESAM& CESAM & CESAM &CESAM & STAREVOL & STAREVOL & STAREVOL\\
Model &  Seismic & SSM &  slow (A)  &  moder. (B)  & slow (A) &  moder. (B) & fast (C) \\
\hline \hline 
X$_{i}$ & 0.7064& 0.7075	&0. 7078	& 0.7117	& 0.7015  &  0.7015
&  0.7015\\
Y$_{i}$  &	0.2722	&	0.2729 &0.2727	 &	0.2690	& 0.280  & 0.280 &
0.280\\
X$_{c}$ &	0.3371 	&	0.361 &0.364	&	0.3522	& 0.3397&   0.3422
& 0.3456\\
Y$_{c}$  &	0.6428 	&    0.618  & 0.615	 &	0.6301	& 0.609  &  0.6376
& 0.6341\\
T$_{c}$  &15.71	&   15.64 & 15.63  &    15.54&	       15.67	& 15.63 & 15.55\\
$\rho_{c}$ &      153.7         & 147.8   &     147.1                   &
146.     &       155.3          &        154.4      &
 153.1     \\
Y$_{s}$ &	0.251  	&0.245	& 0.242 &0.253	 &	0.248 &  0.267 & 0.265\\
$\alpha$	& 2.04 	&	1.77&1.77	&	1.725& 1.7378	 & 1.7378 &
1.7378\\
r$_{BZC}$	& 0.7113      	&	0.714&0.715	 &	0.7241	& 0.719 &  0.7228
& 0.7235\\
(Z/X)$_s$	& 0.0245 &   0.0245  &0.0245		 &   	0.0245 &0.0230	& 0.0250 & 0.0247 \\
\hline 
\end{tabular}
\footnotetext[1]{ The indices i and s are for initial and surface, 
the central temperature $T_C$ is in million degrees, 
boron flux in $\rm 10^6cm^{-2}s^{-1}$ }
\label{tab:sunmod}
\end{table}

\begin{figure}[t]
\begin{center}
\includegraphics[width=12cm]{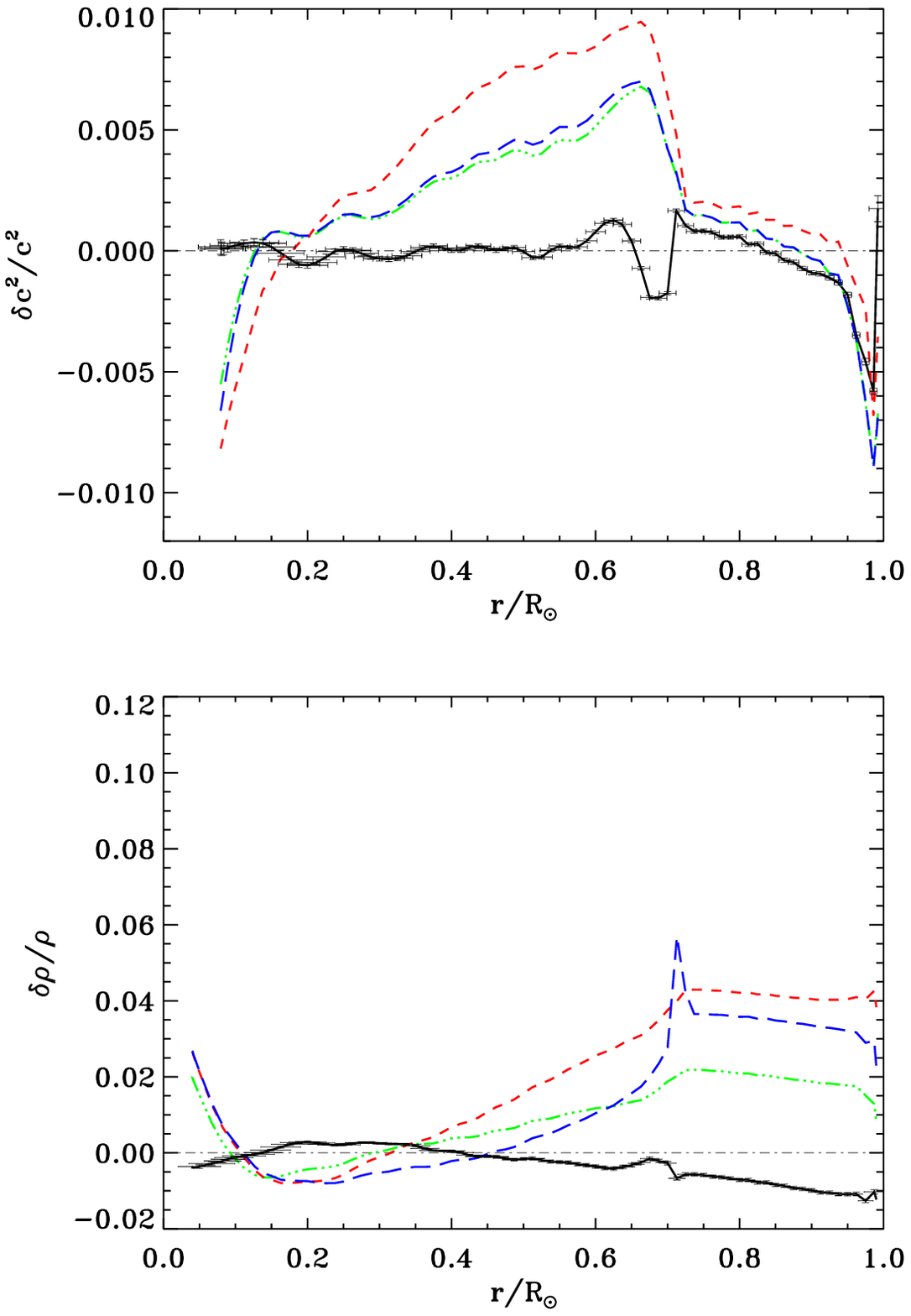}
\caption{Radial squared sound speed and density differences between
observations and models computed with the CESAM code. Seismic model
\citep{Turck2001a,Turck04a} is in black with error bars coming from the
seismic data.  The present models (standard GN composition) without
rotation, with low rotation (model A) or moderate rotation, of 20 km/s
initial model with a magnetic break of the surface at the arrival on the
main sequence (model B) are respectively in green (dotted dashed line),
blue (large dashed line) and red (small dashed line). }
\vspace{-0.5cm} 
\label{fig:soundspeed}
\end{center}
\end{figure}
\subsection{Discussion}

 These detailed calculations are self-consistent and rely on the use of a set of parameters that have been extensively tested throughout the HR diagram. They consider initial rotation generally smaller than used in previous litterature \citep{Chaboyer, Krishnamurthi} and show  slower differential rotations in the core than found in previous works \citep{Pinsonneault} or the most recent
of \cite{CharbonnelTalon05}.
Figure~\ref{fig:sismo} shows a first  comparison between the
different profiles and the presumed observed rotation profile. It is clear
that the models with an initial reasonable rotation velocity + magnetic braking are not supported by
the observations. As indicated by \cite{Garcia2007}, the helioseismic data
reject an increase by a factor 10 for the core rotation.  On the contrary
the slow initial rotation leads to a profile for the present Sun for
which the contrast between the core and the surface angular velocity is
very similar to the presumed observed one. The contrast is similar but of
course none of the different models investigated here create a flat
rotation between 0.25 and 0.7 solar radius. Within the framework of the
present study, models A$_S$ and A$_C$ corresponding to an extremely slow rotating young Sun are more compatible with the central
observations. However, this cannot be taken at face value because this model seems rather irrealistic because it did not contain any magnetic braking (see Figure 2 of  \cite{Bouvier09}.
So one can imagine a slightly greater initial rotation, typically 5 or 10 km/s at the ZAMS, with a central rotation value slightly eroded by some other process. 
The discrepancy
between helioseismology and the profiles presented here calls for the
inclusion of additional dynamical processes that efficiently extract
angular momentum from the radiative interior. This result  confirms the most recent works.  Probably the radiative magnetic field
\citep{Eggenberger05,Yang1, Denissenkov07} or the internal gravity waves
\citep{CharbonnelTalon05} or both, could flatten the profile  but their role could be smaller than thought previously.

\subsection{The sound speed profile, composition and neutrino predictions}

Figure~\ref{fig:soundspeed} compares the squared sound speed profile
difference between seismic observations and the models computed with
CESAM. The used acoustic modes are given in \cite{Couvidat03a}. We present
the three models (standard, low rotation and moderate rotation) computed
with the version of CESAM used in the present study. These models slightly
differ from the standard models already published  due to the large reduction of  the ($^{14}$N, p) reaction rate mentioned in section 2. The CNO combustion is rather small,  but this strong decrease, previously studied in  \cite{Turck2001b}, leads to a negative sound speed difference in the core partly compensated by a small increase in the rest of the radiative zone. These two differences increase the difference with
observations. But what we would like to emphasize in this figure is the intercomparison
between models with and without rotation.

Even if these models must be more representative of the real Sun, the sound
speed profile obtained for the two calibrated CESAM rotating models A and B
are in less agreement with the observed profile than the standard model. In
fact the rotation-induced mixing has reduced the effect of the microscopic
diffusion (see also \S~4) partly inhibited by the macroscopic
motions. As discussed earlier, such effect is practically not visible in
models A, where microscopic diffusivity remains larger than both the
vertical turbulent and the effective diffusivity, but it is clearly visible
in models B and C rotating faster on the ZAMS.

As a natural consequence, Table~\ref{tab:sunmod} shows also that the
central temperature is slightly reduced in rotating models in
comparison with the SSM. For the CESAM models calibrated in radius and
luminosity at 10$^{-4}$ level, the predictions for the neutrinos are
always worse than those of the seismic model (see details in Table 1).
Moreover in CESAM models where we have kept superficial Z/X constant,
one may notice that the surface helium content also increases with the
initial rotation due to turbulent flow just below the convective zone,
as previously mentioned in Section 4 for the STAREVOL results but by a
smaller amount (comparison of the two codes in Table 6).  This will
contribute to reconcile the observed helium content to the predicted
one when one will use the most recent composition of Asplund et al
(2009), if the initial rotation is sufficient to allow turbulent flow
at the base of the convective zone and in the tachocline.
 
\section{CONCLUSION AND PERSPECTIVES}
In this paper we have shown the following facts:

- We have examined  three initial rotation rates (models A, B and C),
choosing initial angular momentum contents corresponding to 2.5 km/s, 20
km/s and 50 km/s at the ZAMS.  The last two  values  have also been adopted in other
studies of the rotating Sun and solar-type stars
\citep{Yang1,Eggenberger05,TalonCharbonnel2005,Chaboyer}.   At 4.6 Gyrs, the three cases show a radial gradient of rotation in the core.
Its amplitude  depends strongly
on the initial rotation rate. If this one is small, the radial gradient
is established during the contraction phase and is
amplified during the subsequent evolution up to the present
Sun.  For models rotating faster, one notes much
steeper gradients at the age of the Sun, the angular momentum losses
associated with magnetic braking on the ZAMS are responsible for the rather small decrease on the ZAMS.

- The transport of angular momentum in the solar radiative zone during the
main sequence appears extremely small  and the
meridional circulation in the radiative zone is smaller by 10 orders
of magnitude in comparison with the observed convective meridional
circulation velocity at 99 \% R$_\odot$. This process is even slower
than the microscopic diffusion (gravitational settling) that we use in
the radiative zone. As a first consequence such implementation in a
stellar evolution code is delicate mainly because one needs to solve
four coupled equations with derivation of quantities that practically
do not vary.  In order to settle our conclusions concerning the form and
order of magnitude of the different quantities associated to the
rotation-induced mixing, we have  confronted models obtained with two different codes,
CESAM and STAREVOL, using distinct numerical approaches and the old composition of Grevesse and Noels. 
We have shown, for models A and B, that they lead
to rather similar results and the same kind of profile for the present
Sun, thus validating the results and numerical approaches.
 Such a very large difference between the meridional velocity in the radiative zone ($10^{-7}- 10^{-6}$ cm/s) and in the convective zone ( m/s) 
would naturally produce some turbulent hydrodynamical layer called and this is an
 interesting result of the present calculation.

- Although the combined effect of meridional circulation and shear-induced
  turbulent associated to rotation is small, this study
  allows us to present radial rotation profiles that can be directly compared to the
  seismically observed one.  This study sustains the idea that the
  Sun was not  a rapid rotator after the contraction phase. The angular velocity
  profile we get for models A is not far from the presumed solar one in the core.
Of course  this model
   should not be considered at face value since it does not take into account any magnetic braking generally observed in young clusters, but it can be useful to estimate the interplay between processes.
  The second model (moderate rotation) is more realistic but shows a greater steep gradient in disagreement with the 
  published detection of gravity modes. So one can imagine that the Sun has been in an intermediate case arriving at 5-10km/s at ZAMS and that its rotation profile would have being eroded by some other process.
 
 - Let us stress however that other dynamical processes known to
 generate efficient transport of angular momentum such as magnetic
 fields and internal gravity waves were not yet included in our
 models. The inclusion of these processes is in the scope of a further
 study, and is expected to significantly affect the choice of the
 preferred model. For instance, internal fossil magnetic field
 produces certainly a very small effect on the solar structure
 \citep{Duez09} but may lead to efficient transport of angular
 momentum that would help flattening the angular velocity profile in
 the radiative zone.  But before we would like to study the activity of the very young Sun. Previous works  show how magnetic field can modify the rotation profile
 \citep{Eggenberger05, Denissenkov07} but the action could be improved by the
 introduction of a more sophisticated field topology which preserves
 the stability of such a field \citep{Duez09b}.  The understanding of the magnetic field during the contraction phase has probably a crucial impact on the radial rotation gradient
  and deserves premature estimate of lithium and beryllium destruction at this stage.

- Other processes may modify the present conclusions. For example in
the present treatment we neglect the fundamental role of the
tachocline which must be established at least since the arrival on the
main sequence. The hydro (or magnetohydrodynamical) nature of such a
region may alter the angular momentum and/or chemicals tranport due to
the internal rotation but we have already noticed that a crude
treatment of this region might slightly amplify the present tendancy
on chemicals gradient and structure effect \citep{Brun2}.  In the
present study, we note that if the initial rotation is accompanied by
an efficient magnetic braking, it generates some turbulent flow at the
base of the convective zone which smoothes the helium profile and
increases its abundance at the surface.

- We note that the impact of the rotation on the solar structure is
rather small. As the transport of angular momentum and chemicals
goes from the radiative zone to the convective zone, it implies a
slight reduction of the central temperature and of the helium
content in the radiative zone. Consequently it increases 
slightly the present discrepancy between model and the observed sound
speed. In fact, one cannot exclude other momentum transport which may
come from the convective zone and play the inverse role
\citep{Garaud08,Gough09}. All these other processes must be included
in stellar evolution codes before getting a proper DSM.  We see
  in this study that the description of the dynamics of the
   PMS phase (especially the related magnetic field
  evolution) will be a crucial issue too. In the comparison between
  CESAM and STAREVOL we have noticed some non negligible difference
  during the contraction phase for models with a moderate (the same
  for high) initial rotation rate, this phase illustrates the
  sensitivity of the rotation gradient to the numerical details and to
  the magnetic braking.  Moreover we show in this study that
  independently of the initial rate, the central rotation value does
  not change by more than 50\% during the main sequence. So the final
  comparison of the DSM to the observations will partly  depend on the
  way we shall treat the contraction phase, the inner corresponding
  magnetic field evolution and the magnetic braking phase. The same
  conclusion was already reached discussing the
  problem of lithium in young stars and in the Sun \citep{Brun2,
    Piau}.

- The sound speed profile is practically unchanged when the slow initial
rotation is assumed.  So even if  this
model is probably more representative of the dynamics of the solar interior
than the standard model, it is still an incomplete model and its predictive
character remains limited. For this reason, and considering the consequent
discrepancy on the sound speed  predicted by the present standard model and
the observed
one, which will be  amplified by the recent CNO composition, we continue to recommend the seismic model for any global predictions
(gravity modes or neutrinos).  We have shown in this paper that the predictions
of the seismic model are in very good agreement with all the neutrino detections
including BOREXINO.

\begin{acknowledgements}
We would like to thank Pierre Morel for his dedicated effort to introduce the dynamical processes in CESAM code, Stephane Mathis and Jean Paul Zahn for very
interesting discussions  and the anonymous referee for helping us to improve the paper.
\end{acknowledgements}

\bibliographystyle{apj}

\end{document}